\documentclass[aps,prb,showpacs,preprintnumbers,twocolumn]{revtex4-2}
\usepackage{amsmath,amssymb}
\usepackage{bm}
\usepackage{tipa}
\usepackage{upgreek}
\usepackage{comment}
\usepackage{mathrsfs}
\usepackage{graphicx}
\usepackage{braket}
\usepackage{mathbbol}
\usepackage{booktabs}
\usepackage{amssymb}
\usepackage{enumitem}
\usepackage{gensymb}
\usepackage[normalem]{ulem}
\usepackage{color}
\usepackage{xcolor}
\usepackage[colorlinks,bookmarks=true,citecolor=blue,linkcolor=red,urlcolor=blue]{hyperref}
\usepackage{hyperref}

\usepackage{tikz-feynman}

\makeatletter
\let\old@makecaption=\@makecaption
\usepackage{subcaption}
\let\@makecaption=\old@makecaption
\makeatother

\usepackage{relsize}

\begin{document}

\title{Chern-Simons Modified-RPA Eliashberg Theory of the $\nu=\frac{1}{2}+ \frac{1}{2}$ Quantum Hall Bilayer}
\author{Tevž Lotrič and Steven H. Simon}

\affiliation{Rudolf Peierls Centre for Theoretical Physics, Parks Road, Oxford, OX1 3PU, UK}

\begin{abstract}
   The $\nu=\frac{1}{2} + \frac{1}{2}$ quantum Hall bilayer has been previsously modeled using Chern-Simons-RPA-Eliashberg (CSRPAE) theory to describe pairing between the two layers. 
However, these approaches are troubled by a number of divergences and ambiguities.   By using a ``modified"  RPA approximation to account for mass renormalization, we can work in a limit 
where the cyclotron frequency is taken to infinity, effectively projecting to a single Landau level.  This, surprisingly, controls the important divergences and removes ambiguities found in prior attempts at CSRPAE.   Examining BCS pairing of composite fermions we find that the angular momentum channel $l=+1$ dominates for all distances $d$ between layers and at all frequency scales.   Examining BCS pairing of composite fermion electrons in one layer with composite fermion {\it holes}  in the opposite layer, we find the $l=0$ pairing channel dominates for all $d$ and all frequencies.  The strength of the pairing in these two different descriptions of the same phase of matter is found to be almost identical.  This agrees well with our understanding that these are two different but dual descriptions of the same phase of matter. 
   \end{abstract}

\maketitle

Quantum Hall bilayers have been a subject of intense investigation since the first experiments on these systems almost thirty years ago (see Refs.~\cite{Experiment_Review,Pinczuk,Eisenstein2004} for reviews).   Conceptually, these are simple systems: a pair of parallel two-dimensional electron gases separated by a distance $d$ placed in a magnetic field at low temperature.   Nonetheless, they show a vast variety of fascinating phenomena.   Perhaps the problem that has attracted the most interest in this field has been the case of the balanced bilayer with Landau level filling fraction $\nu=\frac{1}{2} + \frac{1}{2}$.    In the limit of small distance $d$ between the layers (compared to the magnetic length $\ell_B$) the system forms an exciton condensate\cite{Experiment_Review} (alternately called a quantum Hall ferromagnet\cite{Moon_Review} or the Halperin ``111" state\cite{Halperin111}).    In this state one can think of each electron in one layer being bound to a correlation hole in the opposite layer, hence forming an exciton.   In contrast, in the limit of large $d/\ell_B$ the system can be considered as two independent $\nu=\frac{1}{2}$ quantum Hall systems, which are known to be composite Fermi liquids\cite{HLR,CompositeFermionsHeinonen,CompositeFermionsJain}.

The composite Fermi liquid may be described either with a Jain wavefunction approach\cite{CompositeFermionsJain}, by attaching two Jastrow factors to the position of each electron, or in a Chern-Simons field theory\cite{HLR,simon_MRPA,CompositeFermionsHeinonen} approach, where a singular gauge transformation is made to attach two infinitely thin flux tubes to each fermion.    
One can also consider the state as being described by a Fermi liquid of composite fermion {\it holes}.
That is, one thinks of the holes in a filled Landau level as being the fundamental degrees of freedom, and attaches flux quanta (or Jastrow factors) to these holes.     For clarity if we mean composite fermion holes we will abbreviate them as CH, whereas when we mean conventional composite fermions, where flux quanta, or Jastrow factors, are attached to the original electron coordinate, we will abbreviate this as CE.  The distinction between the CE and CH Fermi liquids, two states that are related to each other by particle-hole conjugation within a single Landau level, is discussed in some detail by Refs.~\cite{AntiCFLiquid,Geraedts2}.   
While the CE and CH trial wavefunctions do not precisely preserve particle-hole symmetry of the half-filled Landau level, which is expected of the ground state in the absence of Landau level mixing in the clean limit, they are numerically exceedingly close to particle-hole symmetric: for $10,11,12$ electrons on a torus, the overlap of the CE wavefunction state\cite{Fremling,Geraedts2} with its particle-hole conjugate (the CH wavefunction) is above 97\%.

The CE and CH approaches to the half-filled Landau level are supplemented by the Dirac composite fermion approach\cite{Son}, which explicitly preserves particle-hole symmetry of the half-filled Landau level. Neglecting this minor distinction that the Dirac approach precisely respects particle-hole symmetry whereas the CE and CH approaches only approximately respect this symmetry, all three approaches (CE, CH, and Dirac) are believed to correctly represent the universal long wavelength physics of the composite Fermi liquid\cite{Mulligan2,CooperHalperin,DMRG_half_filled}.

Returning now to the $\nu=\frac{1}{2} + \frac{1}{2}$ bilayer, although the two limits of large and small distance $d$ between the layers have been quite well understood for some time, the question that has occupied the community for years\cite{Crossover,Liu2019b,Moon_Review,Eisenstein2004,Experiment_Review,Bonesteel,p_wave,Pwave2,Ezawa_2009,ShouCheng,HF1,HF2,HF3,HF4,HF5,HF6,ED1,ED3,ED0,DMRG,Park1,Park2,Simon1,Simon2,Simon3,Kimchi,Sodemann,Milovanovic,Ye1,Ye2,ICCFL,Cipri_thesis,CipriBonesteel,papicThesis,Bosonization1,Bosonization4,Chakravarty,MilovanovicPredin} is what happens for intermediate $d/\ell_B$.    Only recently a clear picture has finally emerged as to the physics of this regime.     Based partially on the Dirac composite fermion picture\cite{Son}, Sodemann et al. \cite{Sodemann} proposed that the two CE Fermi liquids, when weakly interacting with each other, should BCS pair in the $l=+1$  angular momentum channel (chiral p-wave), and they further proposed that this phase of matter is continuously connected to the exciton condensate at $d=0$.  The idea of BCS pairing of CEs in such bilayers was not new\cite{Bonesteel0,Bonesteel,Chakravarty,Simon1,Simon2}, but it was not previously clear that the BCS paired state of CEs could be the same phase of matter as the exciton condensate. 

Inspired by new experiments in bilayers built from graphene\cite{Crossover,Liu2019b}, an alternative picture was recently constructed.   In this picture one imagines condensing BCS pairs made from a CE in one layer bound to a CH of the other layer in the $l=0$ angular momentum channel (s-wave).  This then gives an apparently different picture of a paired state.  What is emphasized in Ref.~\cite{Crossover} (supplement) is that in this picture, in the limit of tightly bound pairs, projection to the lowest Landau level gives precisely the exciton condensate, or Halperin 111 state.   

To test these two pictures of interlayer pairing, Wagner et al.~\cite{Wagner}  (see also Ref.~\cite{QiWagner}) constructed Jain-style\cite{CompositeFermionsJain} trial wavefunctions 
for BCS paired states both for $l=+1$ CE-CE pairing and for $l=0$ CE-CH pairing.   (These constructions were both based on earlier work of M\"oller, Simon, and Rezayi~\cite{Simon1}.) 
Both approaches were found to be extremely accurate for all values of $d$ when compared with exact diagonalizations on small systems (square overlaps  $ \gtrsim .97$ for system sizes of 6+6 electrons on a sphere where the symmetry-reduced Hilbert space is 252 dimensional), and the two approaches were essentially indistinguishable in how well they performed.  We conclude that both approaches are describing the same physics --- although the mapping between the two approaches is nontrivial.   

To try to access the thermodynamic limit, and in order to gain more physical intuition, one can attempt to address the pairing between the two layers analytically.   Very early in the history of the field, Bonesteel, Macdonald and Nayak~\cite{Bonesteel} described the CEs in each layer using the Halperin-Lee-Read (HLR) Chern-Simons field theory\cite{HLR}.  The bosonic ``glue" that pairs the fermions together between the two layers is the Chern-Simons RPA screened Coulomb interaction.  Ref.~\cite{Bonesteel} then used Eliashberg theory to evaluate the pairing instability, the result of which we call Chern-Simons RPA Eliashberg theory (abbreviated CSRPAE).  Although such calculations are plagued with divergences, these authors were  nonetheless able to argue that the system would be unstable to pairing at any finite $d$, although at the level of this calculation all angular momentum channels of pairing are degenerate. 

A more detailed version of this CSRPAE calculation was attempted much later by Isobe and Fu\cite{Fu}  (other versions were attempted by Refs.~\cite{Chakravarty,Mendoza}).    To control infra-red diveregences, Isobe and Fu introduced a wavevector cutoff $q_c$ which is taken to be a very small fraction of the Fermi momentum.  There are two coupling constants that are calculated in this Eliashberg theory: $\lambda_Z(\omega_m)$, the prefactor of the non-anomalous electron self-energy, which in this calculation diverges as $1/q_c$ at any (fermion) Matsubara frequency $\omega_m$  and is independent of the pairing channel, and $\lambda_\phi^{(l)}(\omega_m)$, the prefactor of the anomalous self-energy, which in this calculation diverges as $\log(q_c)$ at any nonzero Matsubara frequency and depends on the pairing channel $l$.  (There are additional, but integrable, divergences as $\omega_m$ goes to zero, which do not need to be regularlized).    Despite these divergences, it was found that the {\it difference} between the coupling constants $\lambda_\phi^{(l)}(\omega_m)$ in different pairing channels $l$ is non-divergent, so that the arbitrary cutoff need not be implemented when comparing different channels to each other, thus suggesting that the arbitrary cutoff may not be problematic.   In particular, the claim of Isobe and Fu was that the pairing angular momemtum channel $l=+1$ is always favored, i.e., $\lambda_\phi^{(l)}(\omega_m)$ is always most negative for $l=+1$.  This pairing channel agreed with earlier trial wavefunction work of M\"oller, Simon, Rezayi\cite{Simon1} as well as with the more recent predictions of Sodemann et al. \cite{Sodemann}.   However, upon repeating this calculation we found that  while $l=+1$ often minimizes $\lambda^{(l)}_\phi$, it can sometimes (depending on $\omega_m$ and $d$) be minimized instead with $l \neq +1$ (see examples of this in  our supplementary material\cite{Supplement}, section \ref{sec:Isobe}), making it hard to draw conclusions confidently as to which pairing channel is actually favored.

Recently the Isobe-Fu calculation\cite{Fu} was generalized by R\"uegg, Chaudhary, and Slager\cite{Slager}
to consider the alternative picture of CEs in one layer and CHs in the other layer.    Again, within CSRPAE theory, the leading divergent terms are independent of pairing channel and one relies on a cutoff to regularize the calculation, although differences in pairing strength are non-divergent allows comparison between different pairings.   The calculation found that the $l=0$ pairing channel  for CE-CH pairing is favored, in agreement with the trial wavefunctions of Ref.~\cite{Wagner}.  In addition, the authors claimed that the CE-CH pairing is stronger than the CE-CE pairing.    
This latter point is a somewhat curious result when compared to the trial wavefunction results of Wagner et al\cite{Wagner} where both  trial wavefunctions seem equivalently good.  
In fact, the comparison made by  R\"uegg, Chaudhary, and Slager\cite{Slager} between CE-CE pairing and CE-CH pairing leaves much unclear because the two approaches have  different divergent terms, so comparison of the coupling strengths depends  on how these divergences are regularized (see Supplementary material\cite{Supplement}, sections \ref{sec:slagerCutoff} and \ref{sec:CECHdivergences}).
While it is always the case that CE-CH pairing is favored compared to CE-CE pairing, i.e., $\lambda_\phi^{(l=0),\textsc{CE-CH}}(\omega_m) \leq \lambda_\phi^{(l=+1),\textsc{CE-CE}}(\omega_m)$, depending on the cutoff and $\omega_m$ this inequality may either be greatly unequal or may be very close to an equality.

The above mentioned Chern-Simons Eliashberg calculations\cite{Bonesteel,Fu,Slager} are excellent starting points for further analytic work which we shall pursue here.  These prior calculations, however, have a number of clear shortcomings:   (1) As mentioned above, the CSRPAE calculations with CE-CE pairing are somewhat ambiguous in which pairing channel is actually favored  (supplementary material\cite{Supplement}, section \ref{sec:Isobe}).
(2) The introduction of an arbitrary infra-red cutoff is somewhat unsatisfying and gives room to doubt that the results are reliable.  (3) The fact that CE-CE and CE-CH pairing have different divergences makes it impossible to compare these two calculations in a cutoff-independent way (supplementary material \cite{Supplement} sections \ref{sec:slagerCutoff} and \ref{sec:CECHdivergences}).      (4) The CSRPAE calculations are based on RPA evaluation of a propagator and RPA is known to have a number of problems --- in particular RPA does not correctly put the low energy physics on the interaction scale and the high energy physics on the cyclotron scale\cite{HLR,SimonHalperinMRPA}.   (5) In making comparison of the CSRPAE approach with the successful trial wavefunctions of Wagner et al.\cite{Wagner}  one may also worry that  the wavefunctions are strictly in the lowest Landau level, whereas Chern-Simons RPA theory is not.    As detailed in Supplementary material\cite{Supplement} section \ref{sec:LLprojection}, this is particularly concerning in the case of CH calculations where one cannot even use ``hole" coordinates as the fundamental degrees of freedom unless the system has a finite Hilbert space dimension, such as when the system is projected to a single Landau level.

The purpose of this paper is to repair the many problems of these previous works and for the first time obtain unambiguous results.   This letter will report our main findings with the calculational details relegated to the Supplement\cite{Supplement}.    Surprisingly, a single new physical ingredient added to the prior calculations can, to a large extent, address {\it all} of the above listed shortcomings.   In this paper we extend the CSRPAE calculations to use a so-called {\it modified} RPA (MRPA) approach developed by Simon and Halperin\cite{SimonHalperinMRPA}, rather than the pure RPA. This scheme, based on Landau Fermi liquid theory, puts the low energy physics on the interaction scale while pushing the cyclotron mode up to the correct frequency so that Kohn's theorem and the f-sum rule are properly satisfied. Setting the cyclotron energy $\omega_c = e B/m_b$ to infinity (i.e., taking the limit of the electron bare band mass $m_b$ going to zero) then should remove any physics of this high energy scale from the problem.  While this is not strictly equivalent to lowest Landau level projection, presumably much of the same physics is included.

We now briefly describe the calculation.   More details are given in the Supplement\cite{Supplement}.   The MRPA scheme\cite{SimonHalperinMRPA,simon_MRPA} accounts for mass renormalization via Landau Fermi theory.   The polarization bubble for noninteracting fermions in zero effective field is calculated with an effective mass $m^*$ which is set by the interaction scale.   To preserve sum rules (stemming from Galilean invariance) we must include a Landau Fermi liquid interaction which amounts to an additional current-current interaction term 
 $A \,  {\bf j} \cdot {\bf j}$  with $A = (m_b - m^*)/(n e^2)$ with $n$ the electron density and  $e$ the electron charge and $\bf j$ the current density.  This current-current interaction term is then treated in RPA along with the Chern-Simons gauge interaction and the Coulomb interaction.  
 The remainder of the Chern-Simons Eliashberg calculation follows that of Refs.~\cite{Slager,Fu} and is detailed in Supplementary material\cite{Supplement} section \ref{sec:detailsMRPA}.   Using MRPA rather than RPA in CSRPAE theory we thus abbreviate as CSMRPAE.   If we set $m^* = m_b$ in CSMRPAE we recover the CSRPAE results of Refs.~\cite{Slager,Fu}.

We now consider CSMRPAE in the limit of $m_b$ going to zero.    This limit is meant to represent projection to a single Landau level, 
although as mentioned in Supplementary material section \ref{sec:LLprojection}, once one makes any sort of mean field approximation, some of the detailed structure of the lowest Landau level is lost, such as its particle-hole symmetry.   Remarkably, in this limit we find that the divergences in coupling constants $\lambda_Z(\omega_m)$ and $\lambda_\phi^{(l)}(\omega_m)$ vanish proportional to $m_b^n$ with $n \geq 1$ for any value of $\omega_m$ such that $0 < \omega_m < {\cal O}(\omega_c)$ (see Supplementary material\cite{Supplement} section \ref{sec:CECHdivergences}, \ref{sub:analysisofdivergence})  for both CE-CE pairing and for CE-CH pairing.  By taking the $m_b \rightarrow 0$ limit, we push $\omega_c$ to infinity, removing all divergences at any finite frequency,  and thus remove the need for an ad-hoc $q_c$ cutoff.  

        \begin{figure}[t]
           \includegraphics[width=2.75in
]{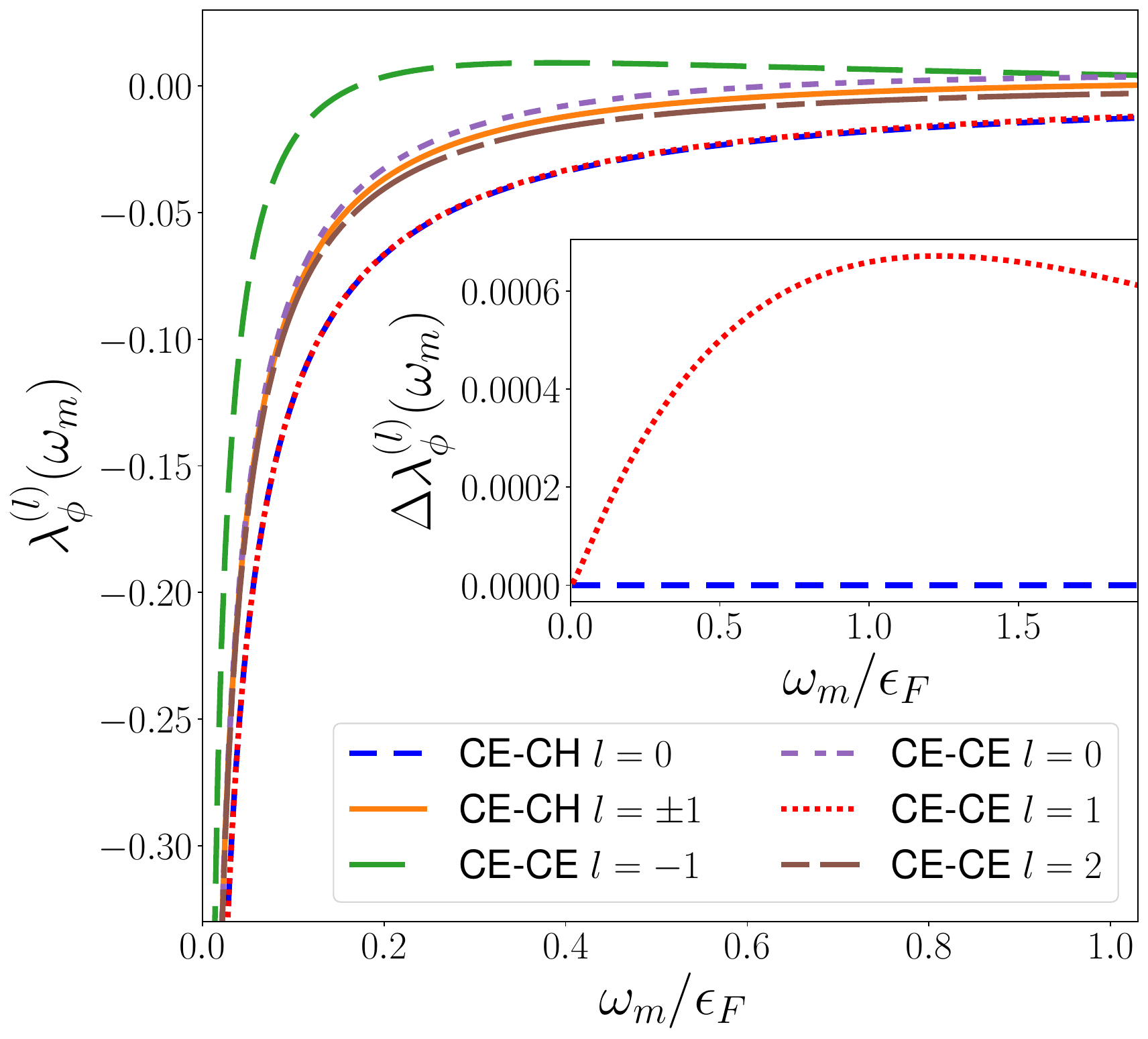}
\caption{The Eliashberg anomalous coupling constant $\lambda_\phi^l(\omega_m)$ calculated using CSMRPAE for interlayer spacing $d/l_B =1$.   The coupling constant is shown for different pairing channels $l$ and for both composite-fermion-electron-composite-fermion-electron pairing (CE-CE) and for composite-fermion-electron-composite-fermion-hole pairing (CE-CH).   We see that CE-CH pairing with $l=0$ and CE-CE pairing with $l=1$ are most attractive (most negative) and are very nearly equal to each other (the two lowest curves almost precisely overlap).     The inset shows a blow up of the difference between the two lowest curves ($l=0$, CE-CH pairing and $l=0$ CE-CE pairing).  The difference is a factor of order 100 smaller than the difference to other pairing channels.  Other values of $d/l_B$, and larger range of $\omega_m$  are shown in the Supplement\cite{Supplement} section \ref{sec:CECEresults_numeric}. }
            \label{fig:lambdacouplings}
        \end{figure}

In this limit, we find that for CE-CE pairing $l=+1$ is now unambiguously the strongest pairing channel for all values of $d$ and $\omega_m$ (see Fig.~\ref{fig:lambdacouplings} and Supplementary material\cite{Supplement} section \ref{sec:CECEresults_numeric}), and for CE-CH pairing $l=0$ remains unambiguously the strongest pairing channel (see Fig.~\ref{fig:lambdacouplings} and Supplementary material\cite{Supplement} section \ref{sec:CECHresults_analytic}).   Further, without the divergences and cutoff dependencies we can now meaningfully compare CE-CE $l=+1$ pairing with CE-CH $l=0$ pairing.   To very high precision (roughly one percent level)  we find that CE-CE pairing and CE-CH pairing are equivalently strong  (see Fig.~\ref{fig:lambdacouplings}, inset).   We also analyze this comparison analytically in Supplementary material\cite{Supplement} sections \ref{sec:CECHresults_analytic} A-C.    The equivalence between the CE-CE and CE-CH pairing is rather surprising given that two very different integrals need to give almost precisely the same result. 
We show further (Supplementary material\cite{Supplement} section \ref{sub:adjustment}) what small modification of our MRPA approximation would make them exactly equal.

We note in passing that, as pointed out in Ref.~\cite{Wagner} (see supplement therein) both CE-CE and CE-CH pairing are equally able to remain paired in the presence of density imbalance between the layers, so long as the total filling remains $\nu_T = 1$, in agreement with experiment\cite{Experiments2,Experiments4}.   We  reiterate this argument in Supplementary material\cite{Supplement} section \ref{sec:imbalance}.  Imbalance will be studied in more depth in a forthcoming work.

To further demonstrate the usefulness of our approach, we now examine a number of extensions.   First, we can consider other filling fractions, see Supplementary material\cite{Supplement} section \ref{sub:otherfilling}.     The $\nu=1/4 + 3/4$ case and the $\nu=1/6 + 5/6$ cases described in the CE-CH picture are unambiguously found to be in the $l=0$ pairing channel.    At $\nu=1/4 + 1/4$ in the CE-CE picture we find that the $l=1$ channel is unambiguously favored.  However, at $\nu=1/6+1/6$, in the CE-CE picture $l=0$ pairing is favored at low frequency (compared to the Fermi energy), but $l=1$ is (only very slightly) favored at higher frequency.  While this leaves a slight ambiguity in the result, it strongly suggests that $l=0$ pairing is realized (this ambiguity was also seen by Ref.~\cite{Fu}, although $l=1$ is more strongly favored at high frequency in that case making the results more ambiguous).   This suggests that $1/6+1/6$ would be interesting to examine further either in numerics or experiment.  Such CE-CE $l=0$  pairing would have zero quantized Hall drag at zero temperature\cite{SenthilMarstonFisher,ReadGreen,KimNayak} as compared to all of the other states considered here which have $h/e^2$ quantized Hall drag resistance.  

We can further examine whether changing the inter-electron interaction might change our results, see Supplementary material\cite{Supplement} section \ref{sub:potentials}.   Assuming that we start with two composite Fermi liquids before we turn on the interlayer interaction, we find that the favored pairing channel is remarkably insensitive to the details of the interelectron interaction within our CSMRPAE approximation.  We have examined (1)  $V(q) \sim 1/(q + q_0)$ which models interaction in the presence of nearby metal screening layers (2)   $V(q) \sim e^{-q^2 w^2}/q$ which models the effects of finite well-width (3) longer ranged potentials $V(q) \sim q^{-2 + \eta}$ with $\eta \in (0,1)$,  (4) Gaussian potentials $V(q) \sim \exp(-q^2 w^2)$, and (5) inclusion of Landau level form factors $V(q) \sim [L_n(q^2 \ell_B^2/2)]^2/q$ with $L$ the Laguerre polynomials.   We use the same form for inter- and intra-layer interaction, although we reduce the strength of the inter-layer compared to intra-layer  (See Supplementary material\cite{Supplement} section \ref{sub:potentials} for the full range of parameters that have been examined).  For $\nu=1/2 + 1/2$ in all cases,  we find $l=1$ favored for CE-CE pairing and $l=0$ favored for CE-CH pairing with the two descriptions being very close to degenerate.  This strongly suggests that very similar physics should occur in a wide range of two dimensional electron systems independent of details. 

Finally we turn to examine the robustness of our results to deformations of the spatial metric.  Such deformations are of particular interest\cite{PhysRevLett.110.206801,KunYangAnisotropy,Yang_tilded_field_neutral,Yang_tilded_field_charged,Papic_Haldane,Papic_Tilted_Field,PhysRevB.86.035122} because some physical two-dimensional electron systems have anisotropic effective mass.   Even with isotropic effective mass, tilted magentic field can make the single-particle orbitals anisotropic.  The simplest case to study is that of a Gaussian inter-electron interaction (case (4) above).    As pointed out by Ref.~\cite{KunYangAnisotropy}, for a system projected to the lowest Landau level, the Gaussian interaction allows one to make a unitary transformation that implements an area preserving diffeomorphism without changing the spectrum --- thus implying complete robustness against geometric deformation.  We discuss geometric deformation further in Supplement \cite{Supplement} section \ref{sec:imbalance_anisotropic}, where we argue that the pairing symmetry remains the same if one examines the system in rescaled coordinates, and  we conjecture that the gap will always be robust to such deformation.

To conclude, we believe our approach of looking at the $m_b \rightarrow 0$ limit of CSMRPAE has satisfactorily tamed the divergences and ambiguities of CSRPAE theory which have been problematic for several decades.   Our  main results for $\nu=1/2 + 1/2$ are: 
 for CE-CE pairing the $l=+1$ pairing channel is unambiguously the strongest, and for CE-CH pairing the $l=0$ pairing channel is unambiguously the strongest.  To very high precision we also find that these two cases pair with the same strength in agreement with the results of prior trial wavefunction calculations\cite{Wagner}.  This is rather satisfying since we believe that the two types of pairing are simply different descriptions of the same physics.  In fact, it is perhaps a bit surprising that our two 
approximate approaches are so  closely equivalent
given that we have 
not enforced any sort of symmetry between the two.  One might think that this near equivalence is a result of particle-hole symmetry (which itself has been broken by the Chern-Simons calculational approach even in the $m_b \rightarrow 0$ limit).  However, even given a perfect particle-hole symmetry, it is not obvious that binding CEs to CEs should be precisely equivalent to binding CEs to CHs.   This should be interpreted as a nontrivial duality which is surprisingly accurately respected by the CSMRPAE approach.
It is an open question whether the two apparently different types of pairing might look more equivalent within the Dirac CF theory, where particle-hole symmetry is manifest at least within each layer.   We comment, however, that the system does not need to have particle-hole symmetry in order for the CSMRPAE to predict the (very near) degeneracy between CE-CE $l=+1$ pairing and CE-CH $l=0$ pairing.    See Supplementary material\cite{Supplement} section X for further elaboration of these issues.

{\it Acknowledgements:}   SHS acknowledges helpful conversations with D. X. Nguyen. TL was supported by the Rudolf Peierls Centre for Theoretical Physics UROP summer internship.  SHS is funded by EPSRC
grant EP/S020527/1. Statement of compliance with EPSRC policy framework on research data:
This publication is theoretical work that does not require supporting research data.

%


\newpage
\clearpage

\onecolumngrid

\begin{center}
    {\large\bf Supplementary Material for:}

\vspace*{10pt}

{Chern-Simons Modified-RPA Eliashberg Theory of the $\nu=\frac{1}{2}+ \frac{1}{2}$ Quantum Hall Bilayer}

\vspace*{10pt}

{Tevž Lotrič and Steven H. Simon}%

\end{center}

\section{Results of Isobe-Fu Calculation}
\label{sec:Isobe}

In this section we reproduce results of the Isobe-Fu calculation\cite{Fu} which are based on RPA.  While that work claimed that the $l=+1$ pairing channel is always favored, we find that their results have some ambiguity.  In particular, at higher frequencies, higher values of angular momenta $l$ can be favored with the transition occurring at lower frequency for larger $d$. A more detailed calculation would be required to confirm the claim of Ref.~\cite{Fu} that $l=+1$ is always the most stable pairing channel. 

\begin{figure}[h]
    \centering
    \begin{subfigure}{0.48\textwidth}
      \centering
      \includegraphics[width=2.5in]{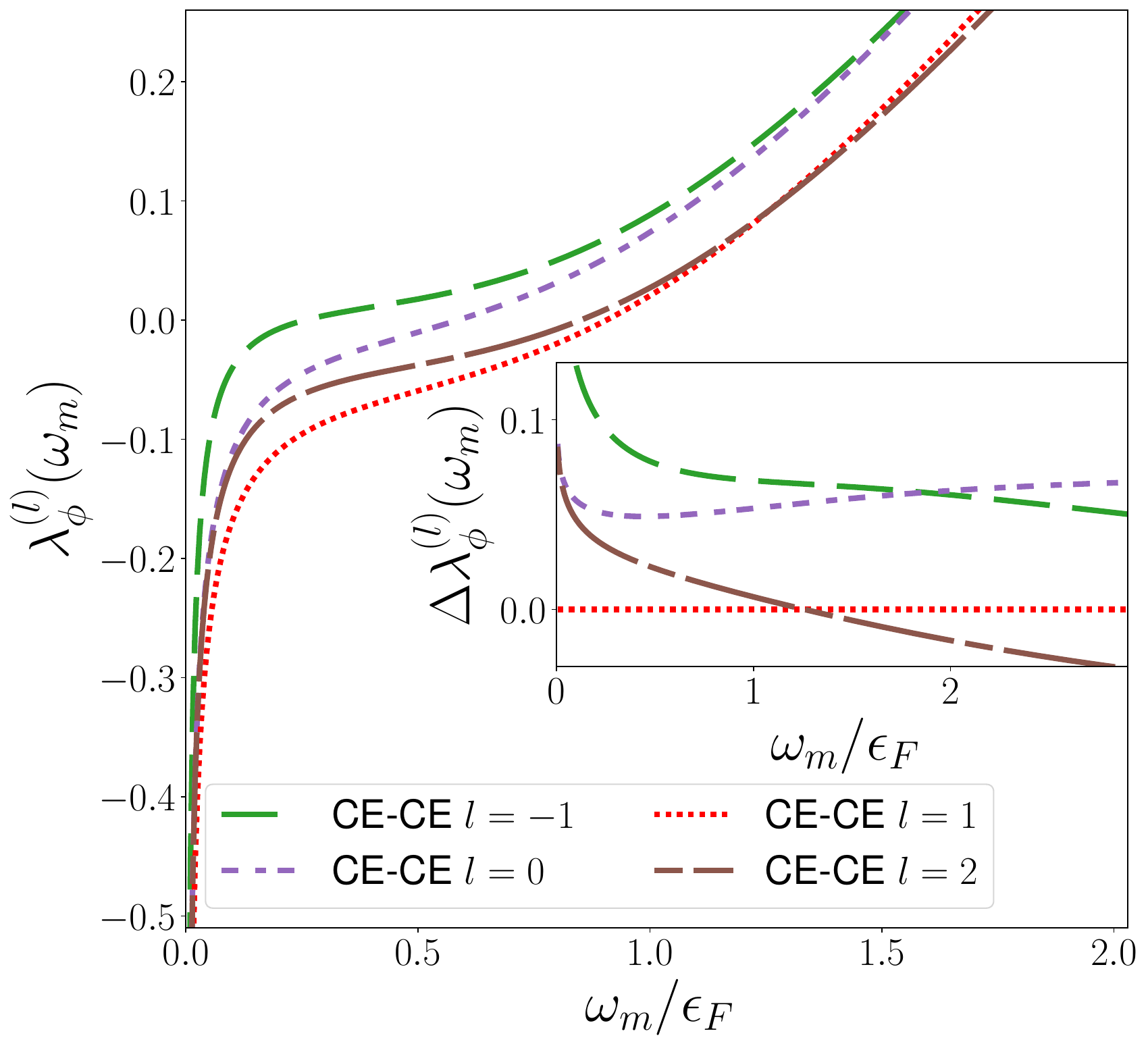}
      \caption{$d/l_B=1$}
      \label{fig:Fu_d_1}
    \end{subfigure}
    \begin{subfigure}{0.48\textwidth}
      \centering
      \includegraphics[width=2.5in]{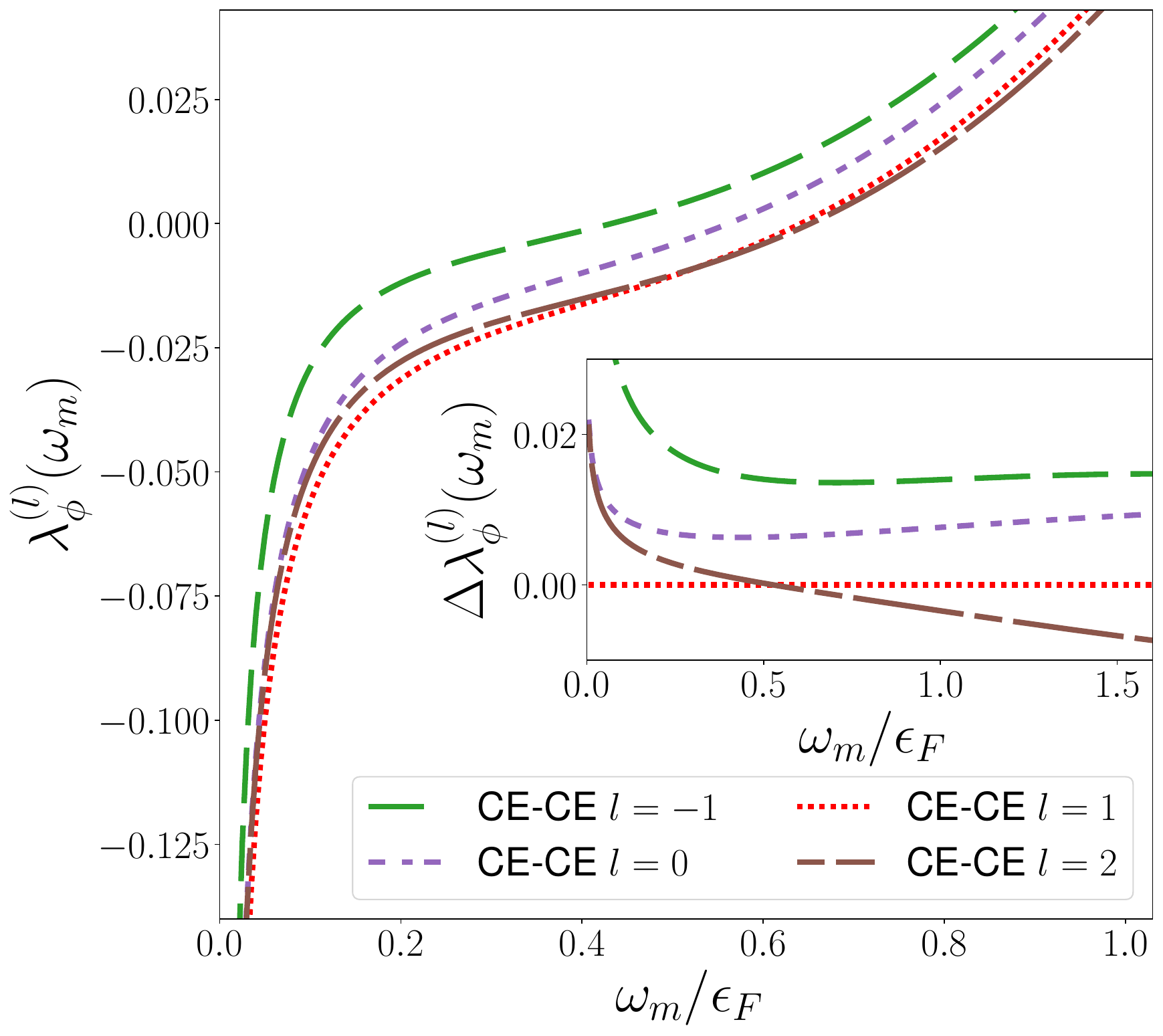}
      \caption{$d/l_B=3$}
      \label{fig:Fu_d_3}
    \end{subfigure}
    \caption{The results for $\lambda_\phi^{(l)}(\omega_m)$ from the Isobe-Fu \cite{Fu} calculation recreated, showing ambiguity in the favoured channel at high frequency. The insets show the behaviour relative to the possibly-favoured $l=+1$ channel. To regularise divergences, we used a cutoff $q_c=10^{-5}\cdot k_F$ in the Eliashberg integrals, Eq.~\ref{eq:lambdaphi}. This matches the cutoff used by Isobe-Fu.}
    \label{fig:Isobe_Fu_ambiguity}
\end{figure}

Fig.~\ref{fig:Fu_d_1} reproduces the results of Isobe-Fu's Figure 2(a) \cite{Fu},  extending the range of $\omega_m$ up to $2\epsilon_F$.   We show the anomalous self-energy coupling constants $\lambda_{\phi}^{(l)}(\omega_m)$ as a function of $\omega_m$ when $d=1/k_F=l_B$ (the definition of this coupling is given in Ref.~\cite{Fu}, but also in Eq.~\ref{eq:lambdaphi} below).  Negative coupling constant corresponds to an attractive interaction. In Fig.~\ref{fig:Fu_d_3} we show coupling constants as a function of $\omega_m$ for $d=3/k_F=3l_B$ instead. In both cases, we see the $l=1$ channel is the strongest at low frequency, before eventually being replaced by $l=2$ at high enough frequency. In fact at very high $\omega_m$, the channels with larger $|l|$ are preferred (although at high frequency $\lambda_{\phi}^{(l)}(\omega_m)>0$, while a negative value is required for attraction).  It is important to note that for larger $d$, the crossing between $l=1$ and $l=2$ occurs at a lower frequency.

While it seems plausible that $l=+1$ is the dominant pairing channel since it is the most attractive at low frequency, without further calculation, strictly speaking, it is ambiguous since higher $|l|$'s can be more attractive at finite frequency.

\section{The ambiguity of comparing CE-CE to CE-CH within RPA}
\label{sec:slagerCutoff}

\begin{figure}[h]
    \centering
    \begin{subfigure}{0.48\textwidth}
      \centering
      \includegraphics[width=2.5in]{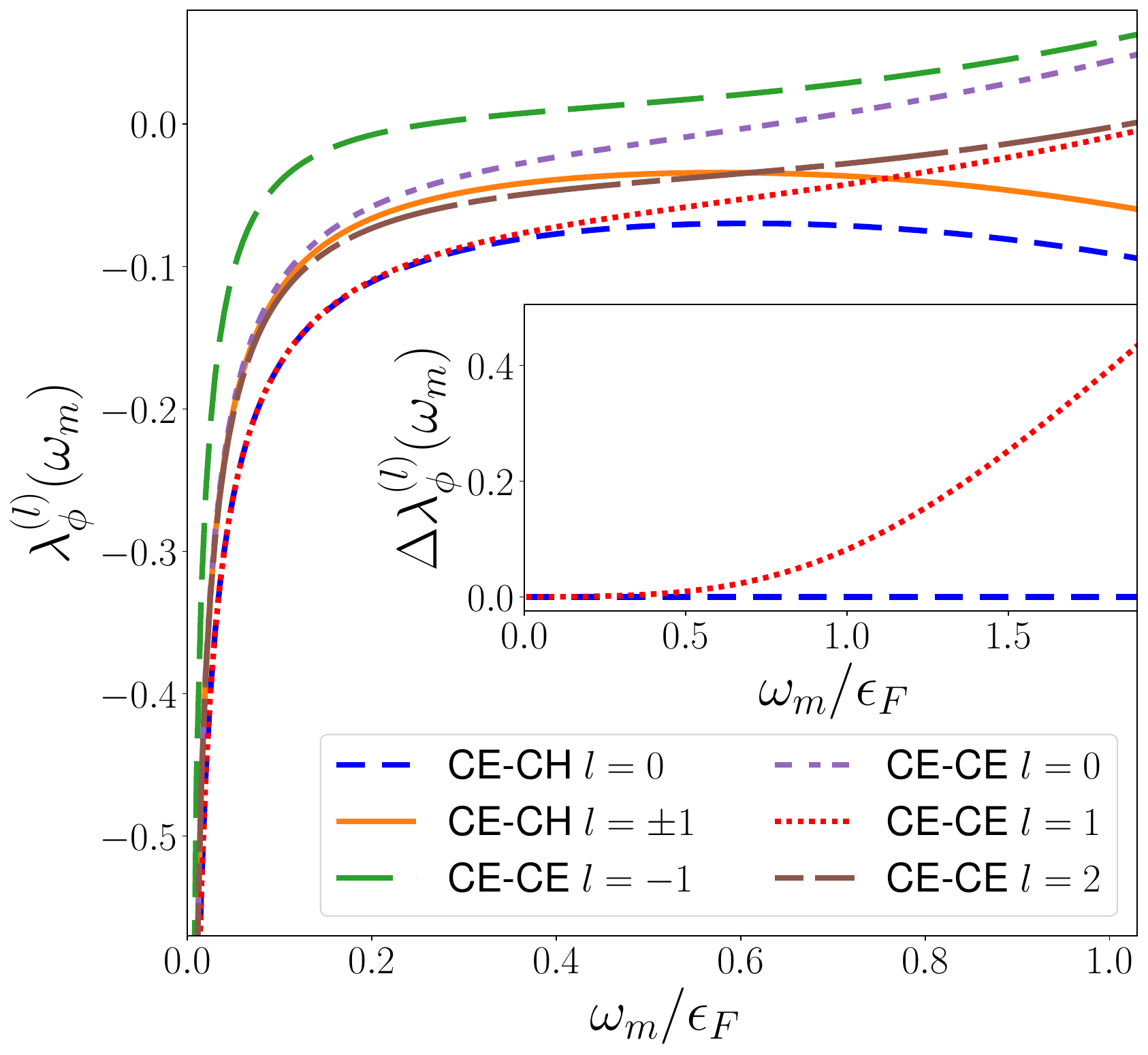}
      \caption{$q_c=10^{-3}\cdot k_F$}
      \label{fig:slager_1e-3}
    \end{subfigure}
    \begin{subfigure}{0.48\textwidth}
      \centering
      \includegraphics[width=2.5in]{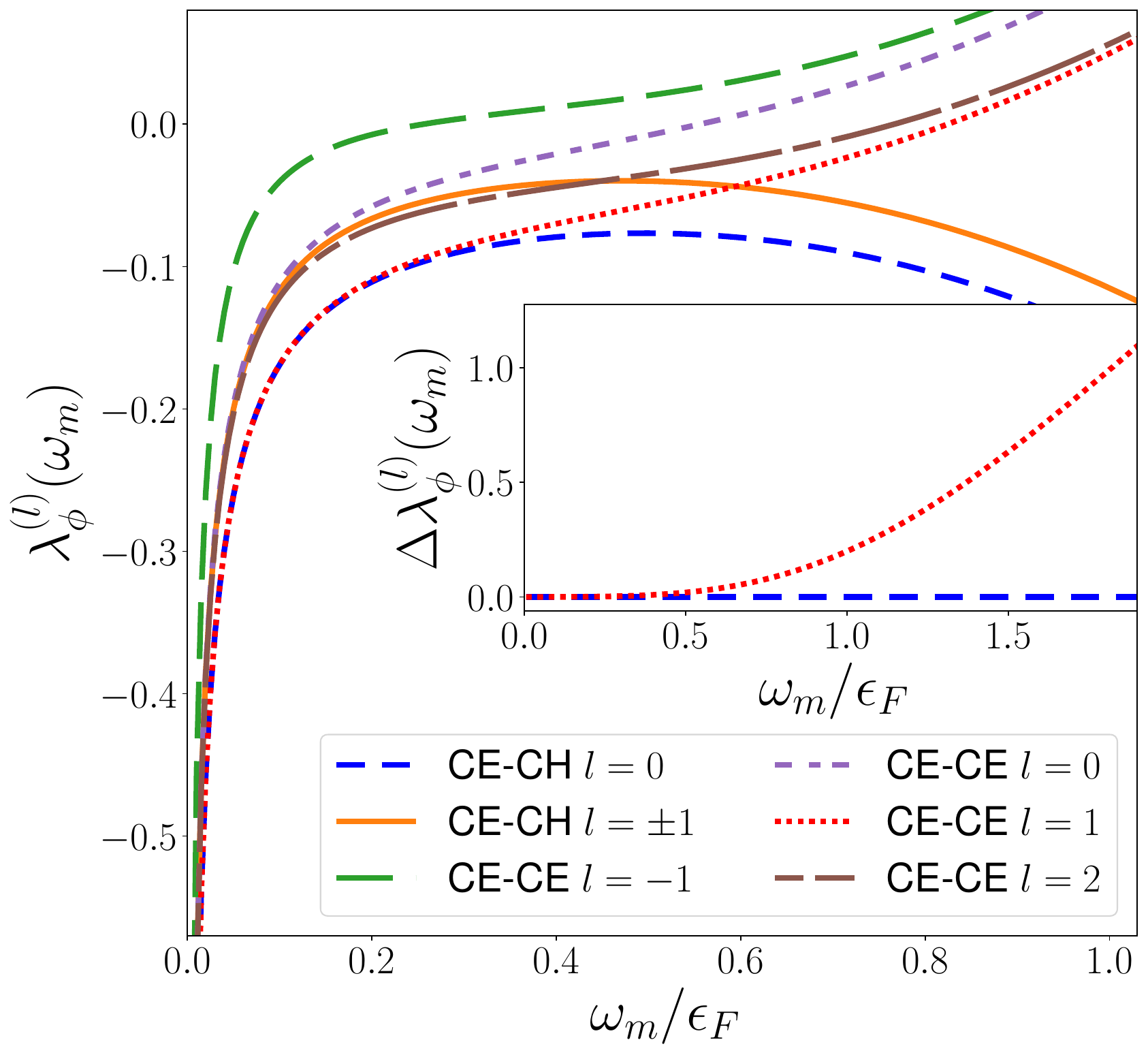}
      \caption{$q_c=10^{-7}\cdot k_F$}
      \label{fig:slager_1e-7}
    \end{subfigure}
    \caption{The different dependence of CE-CE pairing channels on the cutoff relative to the CE-CH channels: at certain frequencies, changing the cutoff can change the ordering of the different channels in pairing strength. The insets show the difference between the two most promising candidates: CE-CE $l=1$ and CE-CH $l=0$. At any fixed $\omega_m$, the difference between the two varies significantly as the cutoff is changed.}
    \label{fig:slager_cutoff_dependance}
\end{figure}

While the introduction of an arbitrary cutoff does not change the ordering of the different CE-CE pairing channels in strength when comparing them to each other, that is no longer the case when comparing CE-CE channels to CE-CH channels. In this section, we present numerical results showing that the comparison between CE-CE and CE-CH channels (as done by \cite{Slager}) significantly depends on the cutoff $q_c$ chosen. Later, in Section \ref{sec:CECHdivergences}, we analytically show that the divergent term for CE-CH coupling is exactly opposite to that for CE-CE coupling. Focusing on Fig.~\ref{fig:slager_cutoff_dependance}, we can clearly see the lines crossing at different frequencies in Fig.~\ref{fig:slager_1e-3} with $q_c=10^{-3}\cdot k_F$, compared to Fig.~\ref{fig:slager_1e-7} with $q_c=10^{-7}\cdot k_F$: for example, at $\omega_m=0.7\epsilon_F$, Fig.~\ref{fig:slager_1e-3} suggests that CE-CE $l=1$ should be favoured over CE-CH $l=\pm1$, while Fig.~\ref{fig:slager_1e-7} suggests CE-CH $l=\pm1$ is the stronger one out of the two. Furthermore, the difference between CE-CH $l=0$ and CE-CE $l=1$ significantly depends on on the cutoff chosen: the insets show that the difference more than doubles when going from $q_c=10^{-3}\cdot k_F$ to $q_c=10^{-7}\cdot k_F$. This makes it hard to interpret the results of \cite{Slager}, where proper care was not taken to handle this divergence.  While the $l=0$ CE-CH pairing is strongest for all the cutoff prescriptions we attempted, whether it wins by a lot or by a little depends significantly on the cutoff.  In particular, Fig.~\ref{fig:slager_cutoff_dependance} shows that most of the difference seen in \cite{Slager} would not be there if a larger value of $q_c$ was chosen.

We note that within the $m_b=0$ MRPA, the low-$q$ divergence is removed, so the results in Fig.~\ref{fig:lambdacouplings} do not rely on any cutoff. As such they are unambiguous in showing which coupling channels are favoured at what frequency and the gaps between channels do not depend on the cutoff chosen.

\section{Details of MRPA Chern-Simons Eliashberg Theory}
\label{sec:detailsMRPA}

Our calculation closely follows that of Isobe and Fu's paper \cite{Fu} and that of R\"uegg, Chaudhary, and Slager\cite{Slager}. In this section, we briefly explain their computation, and show how it is modified to include the MRPA.
\subsection{RPA and MRPA propagators}
We start with the standard HLR Lagrangian\cite{HLR}, adapting it to two layers, adding a Coulomb interaction between the layers, and introducing a charge parameter, $c_s$, in each layer. A layer of CE's has $c_s=1$ and a layer of CH's has $c_s=-1$. In Euclidian form, the action is\cite{Fu,Slager}
\begin{align} \label{eq:euclidian_action}
\begin{split}
    S=\sum_{s\in\{1,2\}}\int\text{d}\tau\text{d}^2\bold{x}\left[\psi_s^\dag(\partial_\tau+ia_0^{(s)}-c_siA_0-\mu_s)\psi_s+\frac{1}{2m^*}\psi^\dag_s\left(i\nabla+\bold{a}^{(s)} -c_s\bold{A}\right)^2\psi_s-\frac{c_si}{4\pi} a_0^{(s)} \hat{z}\cdot\left(\nabla\times\bold{a}^{(s)}\right)\right]\\
    +\frac{1}{2}\sum_{s,s'\in\{1,2\}} c_s c_{s'} \int\text{d}\tau\text{d}^2\bold{x}\text{d}^2\bold{y}\delta\rho_s(\bold{x},\tau)V_{ss'}(\left|\bold{x}-\bold{y}\right|)\delta\rho_{s'}(\bold{y},\tau).
\end{split}
\end{align}
Here, $V_{ss'}(r)=\frac{e^2}{\epsilon\sqrt{r^2+(1-\delta_{ss'})d^2}}$ is the Coulomb interaction between the composite fermions (CFs), and $\delta\rho_s(\bold{x},\tau)=\psi^\dag_s(\bold{x},\tau)\psi_s(\bold{x},\tau)-n_e$ is the CF density fluctuation.

We start with the mean-field solution and account for fluctuations within the RPA. First notice that $a^{(s)}_0$ only appears linearly in $S$, so it acts as a Lagrange multiplier, enforcing
\begin{equation} \label{eq:gauge_curl_electron_density}
    \psi^\dag_s\psi_s=\frac{c_s}{4\pi}\hat{z}\cdot\left(\nabla\times\bold{a}^{(s)}\right)    
\end{equation}
The system is at half-filling, so $\nabla\times\bold{A}=B=\frac{2\pi}{\nu}n_e=4\pi n_e$. The mean field may thus be identified as $\delta\rho_s=0$ and $\bold{a}^{(s)}=c_s\bold{A}$. To expand around the mean field, we let $\bold{a}^{(s)}=c_s\bold{A}+\bold{\Tilde{a}}^{(s)}$ and use Eq.~\ref{eq:gauge_curl_electron_density} on the interaction term in Eq.~\ref{eq:euclidian_action}. This gives us
\begin{equation} \label{eq:mean_field_action}
\begin{split}
    S=\sum_{s\in\{1,2\}}\int\text{d}\tau\text{d}^2\bold{x}\left[\psi_s^\dag(\partial_\tau+i\Tilde{a}_0^{(s)}-\mu_s)\psi_s-i\Tilde{a}_0^{(s)}n_e+\frac{1}{2m^*}\psi^\dag_s\left(i\nabla+\bold{\Tilde{a}}^{(s)}\right)^2\psi_s-\frac{c_si}{4\pi} \Tilde{a}_0^{(s)} \hat{z}\cdot\left(\nabla\times\bold{\Tilde{a}}^{(s)}\right)\right]\\
    +\frac{1}{2}\frac{1}{(4\pi)^2}\sum_{s,s'\in\{1,2\}}\int\text{d}\tau\text{d}^2\bold{x}\text{d}^2\bold{y} \hat{z}\cdot\left[\nabla\times\bold{\Tilde{a}}^{(s)}(\bold{x},\tau)\right]V_{ss'}(\left|\bold{x}-\bold{y}\right|)\hat{z}\cdot\left[\nabla\times\bold{\Tilde{a}}^{(s')}(\bold{y},\tau)\right].
\end{split}
\end{equation}
We move to Fourier space, choosing to continue in the Coulomb gauge for $\bold{\Tilde{a}}^{(s)}$, $\nabla\cdot\bold{\Tilde{a}}^{(s)}=0$ and dropping the tilde from now on. In the Couloumb gauge, we may write $\bold{a}^{(s)}(\bold{q},\tau)=a_1^{(s)}(\bold{q},\tau)(\hat{z}\times\hat{\bold{q}})$. The action in Fourier space is
\begin{align}
    S\left[a_\mu,\psi^\dag_s,\psi\right]&=S_\text{a}\left[a_\mu\right]+S_{\text{af}}\left[a_\mu,\psi^\dag_s,\psi_s\right] \\
    S_\text{a}\left[a_\mu\right]&=\frac{1}{2}\sum_{\omega_m}\int\frac{\text{d}^2q}{(2\pi)^2}\sum_{s,s'\in\{1,2\}}\sum_{\mu,\nu\in\{0,1\}}a_\mu^{(s)}(\bold{q},i\omega_m)\left[\mathcal{D}^{(0)}(q,i\omega_m)\right]^{-1}_{\mu s,\nu s'}a_\nu^{(s')}(-\bold{q},-i\omega_m)  \\
    \left[\mathcal{D}^{(0)}(q,i\omega_m)\right]^{-1}_{\mu s,\nu s'} &= \begin{pmatrix}
        0 & \frac{q}{4\pi}\delta_{ss'} \\
        \frac{q}{4\pi}\delta_{ss'} & -\frac{q^2V_{ss'}(q)}{(4\pi)^2}
    \end{pmatrix}_{\mu\nu}\\
    S_{\text{af}}\left[a_\mu,\psi^\dag_s,\psi_s\right]&=\sum_{s\in\{1,2\}}\sum_{\omega_m,\epsilon_n}\int\frac{\text{d}^2k}{(2\pi)^2}\frac{\text{d}^2q}{(2\pi)^2} \label{eq:af_action}\\
    &\times \psi^\dag_s\left(\bold{q+k},i\epsilon_n+i\omega_m\right)\left[\left(2\pi\right)^2\delta^{(2)}(\bold{q})\delta_{\omega_m,0}\left(-i\epsilon_n+\frac{k^2}{2m^*}-\mu_s\right)+\phi(\bold{q},\bold{k},i\omega_m)\right]\psi_s\left(\bold{k},i\epsilon_n\right) \nonumber\\
    \phi(\bold{q},\bold{k},i\omega_m)&=ia_0^{(s)}(\bold{q},i\omega_m)-\frac{1}{m^*}\left[\hat{z}\cdot\left(\hat{\bold{q}}\times\bold{k}\right)\right]a_1^{(s)}(\bold{q},i\omega_m) \label{eq:phi_matrix}\\
    &+\frac{1}{2m^*}\sum_{\omega_p}\int\frac{\text{d}^2p}{(2\pi)^2}a_1^{(s)}(\bold{p+q},i\omega_m+i\omega_p)\frac{(\bold{p+q})\cdot(-\bold{p})}{|\bold{p+q}||\bold{p}|}a_1^{(s)}(-\bold{p},-i\omega_p) \nonumber
\end{align}
where $\omega_m$ is a bosonic Matsurbara frequency and $\epsilon_n$ is a fermionic Matsurbara frequency. $V_{ss'}(q)=\frac{2\pi e^2}{\epsilon q}e^{-qd(1-\delta_{ss'})}$ is the Fourier transformed Coulomb interaction. The RPA is performed by integrating out the fermionic degrees of freedom, and proceeding to expand the result to second order in the gauge fields $a^{(s)}_\mu(\bold{q},i\omega_m)$. Since the action is quadratic in the fermionic fields, the first step is simple, yielding
\begin{align}
    \mathcal{Z}&=\int D(a_\mu) e^{-S_\text{a}\left[a_\mu\right]}\int D(\psi^\dag,\psi) e^{-S_\text{af}\left[a_\mu,\psi^\dag,\psi\right]}=\int D(a_\mu) e^{-S_\text{a}\left[a_\mu\right]}\det\left[-\hat{G}_0^{-1}+\hat{\phi}\right]=\int D(a_\mu) e^{-S_\text{a}\left[a_\mu\right]-\Delta S\left[a_\mu\right]} \\
    \Delta S &= - \ln\det\left[-\hat{G}_0^{-1}+\hat{\phi}\right]=-\text{tr}\ln\left[-\hat{G}_0^{-1}+\hat{\phi}\right]=-\ln\text{tr}\left[-\hat{G}_0^{-1}\right]+\text{tr}\left[\hat{G}_0\hat{\phi}\right]+\frac{1}{2}\text{tr}\left[\hat{G}_0\hat{\phi}\hat{G}_0\hat{\phi}\right]+\ldots
\end{align}
Here, $\hat{G}_0$ is an operator diagonal in momentum and frequency, whose diagonal elements are the fermionic propagators, $G(\bold{q},i\epsilon_n)=\frac{1}{i\epsilon_n-\xi_\bold{q}}$, with $\xi_\bold{q}=\frac{q^2}{2m^*}-\mu$. We assumed equal chemical potential in both layers, $\mu_1=\mu_2=\mu$. The operator $\hat{\phi}$ is not diagonal in the momenta: the amplitude for a momentum transfer $\bold{q}$ and energy transfer $i\omega_m$ from an initial state with momentum $\bold{k}$ is described by Eq.~\ref{eq:phi_matrix}. Continuing the RPA scheme by expanding to second order in the gauge fields and summing over momenta, which accounts for the polarization bubble, yields the correction
\begin{align}
    \Delta S &=-\frac{1}{2}\sum_{\omega_m}\int\frac{\text{d}^2q}{(2\pi)^2}\sum_{s,s'\in\{1,2\}}\sum_{\mu,\nu\in\{0,1\}}a_\mu^{(s)}(\bold{q},i\omega_m)\Pi_{\mu\nu}(\bold{q},i\omega_m)\delta_{ss'}a_\nu^{(s')}(-\bold{q},-i\omega_m),  \\
    \Pi_{00}(\bold{q},i\omega_m)&=T\sum_{\epsilon_n}\int\frac{\text{d}^2k}{(2\pi)}G(\bold{k+q},i\epsilon_n+i\omega_m)G(\bold{k},i\epsilon_n)=-\int\frac{\text{d}^2k}{(2\pi)^2}\frac{f(\xi_\bold{k+q})-f(\xi_\bold{k})}{i\omega_m-\xi_\bold{k+q}+\xi_\bold{k}}, \label{eq:PI00} \\
    \Pi_{01}(\bold{q},i\omega_m)&= \Pi_{10}(\bold{q},i\omega_m)=0,\\
    \Pi_{11}(\bold{q},i\omega_m)&=\frac{T}{m^*}\sum_{\epsilon_n}\int\frac{\text{d}^2k}{(2\pi)}G(\bold{k},i\epsilon_n)-T\sum_{\epsilon_n}\int\frac{\text{d}^2k}{(2\pi)}\frac{\hat{\bold{q}}\times\bold{k}}{m^*}\cdot\frac{(-\hat{\bold{q}})\times\bold{k}}{m^*}G(\bold{k+q},i\epsilon_n+i\omega_m)G(\bold{k},i\epsilon_n)= \nonumber\\
    &=\frac{n_e}{m^*}-\int\frac{\text{d}^2k}{(2\pi)^2}\left(\frac{\hat{\bold{q}}\times\bold{k}}{m^*}\right)^2\frac{f(\xi_\bold{k+q})-f(\xi_\bold{k})}{i\omega_m-\xi_\bold{k+q}+\xi_\bold{k}},\label{eq:PI11}
\end{align}
with $f(\xi)=\frac{1}{\exp(\xi/T)+1}$ the Fermi distribution function.  Note that in principle, if a superconducting gap opens in the spectrum, the polarization bubble needs to be re-computed accounting for this gap.    If we work close to the superconducting transition, this can be ignored.  
The integrals in Eqs.~\ref{eq:PI00}--\ref{eq:PI11} are analytically evaluated in the zero-temperature case by Isobe and Fu \cite{Fu}, for example. In what follows, $T=0$ is assumed and the Isobe-Fu result is used for the polarization function. While not immediately obvious from the definition Eqs.~\ref{eq:PI00}--\ref{eq:PI11}, $\Pi_{\mu\nu}(\bold{q},i\omega_m)$ is rotationally invariant, depending only on the magnitude of $\bold{q}$. With this, the effective action for the gauge field within the RPA is
\begin{align} 
    S_\text{eff}\left[a_\mu\right]&=\frac{1}{2}\sum_{\omega_m}\int\frac{\text{d}^2q}{(2\pi)^2}\sum_{s,s'\in\{1,2\}}\sum_{\mu,\nu\in\{0,1\}}a_\mu^{(s)}(\bold{q},i\omega_m)\left[\mathcal{D}^\text{RPA}(q,i\omega_m)\right]^{-1}_{\mu s,\nu s'}a_\nu^{(s')}(-\bold{q},-i\omega_m),  \\
    \left[\mathcal{D}^{\text{RPA}}(q,i\omega_m)\right]^{-1}_{\mu s,\nu s'}&=\left[\mathcal{D}^{(0)}(q,i\omega_m)\right]^{-1}_{\mu s,\nu s'}-\Pi_{\mu\nu}(q,i\omega_m)\delta_{ss'}. \label{eq:D_RPA_def}
\end{align}
To implement the MRPA correction on top of this, we first need to find the RPA fermion polarization function \cite{simon_MRPA},
\begin{equation} \label{eq:K_RPA_DEF}
    \left[K^\text{RPA}(q,i\omega_m)\right]^{-1}_{\mu s,\nu s'}=\left[\Pi(q,i\omega_m)\right]^{-1}_{\mu\nu}\delta_{ss'}-\mathcal{D}^{(0)}_{\mu s,\nu s'}(q,i\omega_m).
\end{equation}
The MRPA prescription is to replace $K^\text{RPA}$ by $K^\text{MRPA}$, which is defined in imaginary time as
\begin{align}
    \left[K^\text{MRPA}(q,i\omega_m)\right]^{-1}_{\mu s,\nu s'}&=\left[K^\text{RPA}(q,i\omega_m)\right]^{-1}_{\mu s,\nu s'}+\mathcal{F}_{1,\mu\nu}(q,i\omega_m)\delta_{ss'} \label{eq:K_MRPA_def}\\
    \mathcal{F}_{1,\mu\nu}&=\frac{m^*-m_b}{e^2 n_e}\begin{pmatrix}
        \frac{\omega_m^2}{q^2} & 0 \\
        0 & -1
    \end{pmatrix}_{\mu\nu} \label{eq:MRPA_F1_def}
\end{align}
where $K^\text{RPA}$ is calculated with the effective mass $m^*$.  Here, we make a distinction between the interaction-renormalized effective mass $m^*$ and the bare band mass $m_b$.    The effective mass is set by the interaction scale so that $\hbar^2/m^* = C e^2 \ell_B/\epsilon$ with $\epsilon$ the dielectric constant and $C$ a constant estimated\cite{HLR} to be roughly $C=0.3$.    The MRPA is designed to satisfy Kohn's theorem and the f-sum rule, which are defined in terms of the bare mass $m_b$ rather than the effective mass \cite{simon_MRPA}. From Eq.~\ref{eq:D_RPA_def} and Eq.~\ref{eq:K_RPA_DEF}, we may infer the relationship between $K^\text{RPA}$ and $\mathcal{D}^\text{RPA}$ (sum over repeated indices implied -- Greek alphabet letters are indices of gauge field components, Latin alphabet letters are layer indices)
\begin{equation} \label{eq:D_rpa_from_K_rpa}
    \mathcal{D}^\text{RPA}_{\mu s,\nu s'}(q,i\omega_m)=\mathcal{D}^{(0)}_{\mu s,\nu s'}(q,i\omega_m)+\mathcal{D}^{(0)}_{\mu s,\sigma t}(q,i\omega_m) K^\text{RPA}_{\sigma t,\rho u}(q,i\omega_m)\mathcal{D}^{(0)}_{\rho u,\nu s'}(q,i\omega_m).
\end{equation}
To get the MRPA gauge propagator, $\mathcal{D}^\text{MRPA}$, we simply replace $K^\text{RPA}$ in Eq.~\ref{eq:D_rpa_from_K_rpa} with $K^\text{MRPA}$.

Since $\mu, \nu \in \{0,1\}$ and $s,s' \in \{1,2\}$, the propagators are 4x4 matrices. From here on, we assume that at least one layer is taken to be CE's, so we may set $c_2=1$ in Eq.~\ref{eq:euclidian_action}. We define the in-phase and out-of-phase components of the gauge field as 
\begin{equation} \label{eq:gauge_transformation}
    \begin{split}
        a^+_0=\frac{1}{\sqrt{2}}\left(c_1a^{(1)}_0+a^{(2)}_0\right), &\hspace*{10px} a^+_1=\frac{1}{\sqrt{2}}\left(a^{(1)}_1+a^{(2)}_1\right), \\
        a^-_0=\frac{1}{\sqrt{2}}\left(c_1a^{(1)}_0-a^{(2)}_0\right), &\hspace*{10px} a^-_1=\frac{1}{\sqrt{2}}\left(a^{(1)}_1-a^{(2)}_1\right). 
    \end{split}
\end{equation}
With these components, both the gauge propagators and fermion polarization matrices are block diagonal. In this form, the bare gauge propagator is 
\begin{align}
        \mathcal{D}^{(0)}(q,i\omega_m)^{-1}&=\begin{pmatrix}
            \mathcal{D}^{(0)}_{+,\mu\nu}(q,i\omega_m)^{-1} & 0 \\
            0 & \mathcal{D}^{(0)}_{-,\mu\nu}(q,i\omega_m)^{-1}
        \end{pmatrix}_{\mu\nu}, \\
        \mathcal{D}^{(0)}_{\pm,\mu\nu}(q,i\omega_m)^{-1}&=\begin{pmatrix}
            0 & \frac{q}{4\pi} \\
            \frac{q}{4\pi} & -\frac{q^2}{(4\pi)^2}\left(V_{11}(q)\pm V_{12}(q)\right)
        \end{pmatrix}_{\mu\nu}. \label{eq:plusMinusGaugeForm}
\end{align}
At RPA and MRPA level we then have 
\begin{align}
    \text{RPA:} & \hspace{10px} \mathcal{D}^\text{RPA}_{\pm,\mu\nu}(q,i\omega_m)=\mathcal{D}^{(0)}_{\pm,\mu\nu}(q,i\omega_m)+\mathcal{D}^{(0)}_{\pm,\mu\sigma}(q,i\omega_m)\left(\left(\Pi(q,i\omega_m)\right)^{-1}-\mathcal{D}^{(0)}_\pm(q,i\omega_m)\right)^{-1}_{\sigma\rho} \mathcal{D}^{(0)}_{\pm,\rho\nu}(q,i\omega_m), \\
    \begin{split} \label{eq:MRPA_gauge_whole_expression}
    \text{MRPA:} & \hspace{10px} \mathcal{D}^\text{MRPA}_{\pm,\mu\nu}(q,i\omega_m)=\\
    & =\mathcal{D}^{(0)}_{\pm,\mu\nu}(q,i\omega_m)+\mathcal{D}^{(0)}_{\pm,\mu\sigma}(q,i\omega_m)\left(\left(\Pi(q,i\omega_m)\right)^{-1}-\mathcal{D}^{(0)}_\pm(q,i\omega_m)+\mathcal{F}_1(q,i\omega_m)\right)^{-1}_{\sigma\rho} \mathcal{D}^{(0)}_{\pm,\rho\nu}(q,i\omega_m).
    \end{split}
\end{align}
In order to convert back to the original $a^{(s)}$ basis, we may use the identities
\begin{align}
    \mathcal{D}_{\pm,\mu\nu}&=\langle a_\mu^{\pm}(q,i\omega_m)a_\nu^{\pm}(-q,-i\omega_m) \rangle, \\
    \mathcal{D}_{\mu s,\nu s'}&=\langle a_\mu^{(s)}(q,i\omega_m)a_\nu^{(s')}(-q,-i\omega_m) \rangle, \label{eq:gauge_propagator}
\end{align}
along with the transformation Eq.~\ref{eq:gauge_transformation}. We note that in block-diagonal form, the propagators are identical between the CE-CE ($c_1=1$) and CE-CH ($c_1=-1$) systems. The physics remains different however due to the different transformations Eq.~\ref{eq:gauge_transformation} from the block-diagonal form into the physical layer basis.
\subsection{The effective interaction}
The gauge field mediates interaction between the fermions. Looking at the form of interactions between $a$ and $\psi$ in Eq.~\ref{eq:af_action} and Eq.~\ref{eq:phi_matrix}, we can see that the action includes an effective interaction term
\begin{align}
    \mathcal{V}=-\frac{1}{2}\int_{\bold{k,k',q}}\sum_{s,s'\in\{1,2\}}\sum_{\mu,\nu\in\{0,1\}}M_{\mu\nu}(\bold{k},\bold{k}',\hat{\bold{q}})&\mathcal{D}^\text{(M)RPA}_{\mu s,\nu s'}(\bold{q},i\omega_m)\\ 
    &\times\psi^\dag_s(\bold{k+q},i\epsilon_n+i\omega_m)\psi_s(\bold{k},i\epsilon_n)\psi^\dag_{s'}(\bold{k}'+\bold{q},i\epsilon_n-i\omega_m)\psi_{s'}(\bold{k}',i\epsilon_n),\nonumber\\
    M_{\mu\nu}(\bold{k},\bold{k}',\hat{\bold{q}})&=\frac{1}{2}\begin{pmatrix}
        1 & -i\frac{\hat{z}\cdot(\hat{\bold{q}}\times\bold{k}')}{m^*} \\
        i\frac{\hat{z}\cdot(\hat{\bold{q}}\times\bold{k})}{m^*} & \frac{(\hat{\bold{q}}\times\bold{k})\cdot(\hat{\bold{q}}\times\bold{k}')}{m^{*2}}
    \end{pmatrix}.
\end{align}
The form of these terms may be derived by computing the expectation value of $e^{-S}$ for the action Eq. \ref{eq:euclidian_action}, using the MRPA value of the gauge field propagator, Eqs.~\ref{eq:MRPA_gauge_whole_expression}--\ref{eq:gauge_propagator}. This step follows Isobe-Fu \cite{Fu} exactly.

\subsection{Eliashberg theory}
Following \cite{Fu} we write the Eliashberg equations for the quasiparticle residue $Z_m = Z(i \epsilon_m)$ and the anomolous self-energy $\phi_m^{(l)}$  in the angular momentum $l$ pairing channel as
\begin{eqnarray}
    (1 - Z_n)  \epsilon_n &=& - T \sum_{m}  \int \frac{d^2 q}{(2 \pi)^2} \frac{Z_{n+m} (\epsilon_n + \omega_m)}{Z_{n+m}^2 (\epsilon_n + \omega_m)^2 + \xi_{\bf k + q}^2 + |\phi_{n+m}^{(l)}|^2  }  V_\text{ex}({\bf k}, {\bf q}, i \omega_m)  \label{eq:Eliashberg01} \\ 
 \phi_n^{(l)} e^{i l \theta_{\bf k}
 } &=& - T \sum_{m}  \int \frac{d^2 q}{(2 \pi)^2} \frac{\phi_{n+m}^{(l)} e^{i l \theta_{\bf k+q}}}{Z_{n+m}^2 (\epsilon_n + \omega_m)^2 + \xi_{\bf k + q}^2 + |\phi_{n+m}^{(l)}|^2  }  V_\text{c}({\bf k}, {\bf q}, i \omega_m) \label{eq:Eliashberg02} 
\end{eqnarray}
where
\begin{align}
 V_\text{ex}(\bold{k},\bold{q},i\omega_m)&=-\sum_{\mu,\nu}M_{\mu\nu}(\bold{k},\bold{k+q},\hat{\bold{q}})\sum_{s\in\{1,2\}}\mathcal{D}_{\mu s,\nu s}(q,i\omega),\\
    V_\text{c}(\bold{k},\bold{q},i\omega_m)&=-\sum_{\mu,\nu}M_{\mu\nu}(\bold{k},\bold{-k-q},\hat{\bold{q}})\sum_{s\in\{1,2\}}\mathcal{D}_{\mu s,\nu \overline{s}}(q,i\omega) \label{eq:cooper_potential_def},
\end{align}
with $\overline{s}$ the opposite layer to $s$.   Here 
$\epsilon_n = (2n +1)\pi T$ is a fermionic Matsubara frequency and $\omega_m = 2 m \pi T$ is a bosonic Matsubara frequency. 

Eqs.~\ref{eq:Eliashberg01}--\ref{eq:Eliashberg02} are similar to the Eliashberg equations for typical phonon driven superconductivity\cite{GeneralEliashberg}.   In that case there is usually a natural small parameter which is the phonon energy scale divided by the Fermi energy.   Given this small parameter one can make the approximation that 
\begin{equation}
    Z_{n+m}^2 (\epsilon_n + \omega_m)^2+ |\phi_{n+m}^{(l)}|^2 \ll  \xi_{\bf k + q}^2  \label{eq:inequality}
\end{equation}
and the denominator becomes a delta function $\delta(\xi_{\bf q + k})$ restricting the $q$ integral to the Fermi surface.    This assumption is also made by \cite{Fu,Bonesteel,Slager,Mendoza,Chakravarty} for the quantum Hall bilayer, although it is harder to justify.   As we will discuss further in section \ref{sub:analysisofdivergence} below, this is assumption is rather dangerous in the quantum Hall bilayer case and can give some spurious results and it is important to be careful to distinguish which results are robust and which results are a property of the approximation.   Note that Ref.~\cite{Chakravarty} does correctly point out the danger of this assumption but they make the assumption nonetheless, as will we for much of this work (however, see the more detailed discussion of section \ref{sub:analysisofdivergence}).

At this stage we will still follow \cite{Fu,Bonesteel,Slager,Mendoza,Chakravarty} and make the assumption Eq.~\ref{eq:inequality}, reconsidering it only in section \ref{sub:analysisofdivergence}.   Thus we obtain
\begin{eqnarray}
    (1 - Z_n)  \epsilon_n &=& -\pi T \sum_{m} \frac{Z_{n+m} (\epsilon_n + \omega_m)}{\sqrt{Z_{n+m}^2 (\epsilon_n + \omega_m)^2  + |\phi_{n+m}^{(l)}|^2}  }  \lambda_{Z}(\omega_m), \\ 
 \phi_n^{(l)} e^{i l \theta_{\bf k}
 } &=& - \pi T \sum_{m}  \frac{\phi_{n+m}^{(l)} e^{i l \theta_{\bf k+q}}}{\sqrt{Z_{n+m}^2 (\epsilon_n + \omega_m)^2 + + |\phi_{n+m}^{(l)}|^2 } } \lambda_{\phi}^{(l)}(\omega_m)
\end{eqnarray}
where the coupling constants $\lambda_{Z}(\omega_m)$ and $\lambda_{\phi}^{(l)}(\omega_m)$ are given by
\begin{align}
    \lambda_{Z}(\omega_m)&=\int \frac{\text{d}^2q}{(2\pi)^2}\delta(\xi_\bold{k+q})V_\text{ex}(\bold{k},\bold{q},i\omega_m),  \label{eq:lambdaz} \\
    \lambda_{\phi}^{(l)}(\omega_m)&=\int\frac{\text{d}^2q}{(2\pi)^2}\delta(\xi_\bold{k+q})V_\text{c}(\bold{k},\bold{q},i\omega_m)\left(1+\frac{q}{k_F}e^{il\left(\theta_\bold{q}- \theta_\bold{k}\right)}\right)^l. \label{eq:lambdaphi}
\end{align}
The angular parts of these integrals may be performed analytically, to yield \cite{Fu}, \cite{Slager}
\begin{align}
    \begin{split} \label{eq:eliashberg_Z}
    \lambda_{Z}(\omega_m)=\frac{1}{(2\pi)^2}\frac{m^*}{k_F}\int_0^{2k_F}\text{d}q\left\{-\frac{1}{\sqrt{1-\left(\frac{q}{2k_F}\right)^2}}\left[\mathcal{D}_{+,00}(q,i\omega_m)+\mathcal{D}_{-,00}(q,i\omega_m)\right]\right. \\
        \left.-\frac{k_F^2}{m^{*2}}\sqrt{1-\left(\frac{q}{2k_F}\right)^2}\left[\mathcal{D}_{+,11}(q,i\omega_m)+\mathcal{D}_{-,11}(q,i\omega_m)\right]\right\}
    \end{split}
    \\
    \begin{split} \label{eq:eliashberg_phi}
    \lambda_{\phi}^{(l)}(\omega_m)=\frac{1}{(2\pi)^2}\frac{m^*}{k_F}\int_0^{2k_F}\left\{\frac{1}{\sqrt{1-\left(\frac{q}{2k_F}\right)^2}} \cos\left(2l\sin^{-1}\frac{q}{2k_F}\right)c_1\left[\mathcal{D}_{+,00}(q,i\omega_m)-\mathcal{D}_{-,00}(q,i\omega_m)\right]\right.\\
    +\frac{k_F}{m^*}\sin\left(2l\sin^{-1}\frac{q}{2k_F}\right)(1+c_1)\left[\mathcal{D}_{+,01}(q,i\omega_m)-\mathcal{D}_{-,01}(q,i\omega_m)\right]\\
    \left.-\frac{k_F^2}{m^{*2}}\sqrt{1-\left(\frac{q}{2k_F}\right)^2}\cos\left(2l\sin^{-1}\frac{q}{2k_F}\right)\left[\mathcal{D}_{+,11}(q,i\omega_m)-\mathcal{D}_{-,11}(q,i\omega_m)\right]\right\}.
    \end{split}    \end{align}
In our calculations, we use the MRPA result for the propagator. In the CE-CH case, $c_1=-1$, the second term in Eq.~\ref{eq:eliashberg_phi} is zero, so as mentioned in \cite{Slager} for the RPA, the CE-CH pairing for $l$ and $-l$ is of equal strength, and this is also true at MRPA level. We note that $\lambda_{Z}(\omega_m)$ does not depend on $c_1$, so it will be exactly equal for CE-CE and CE-CH couplings: only $\lambda_{\phi}^{(l)}(\omega_m)$ changes based on which coupling is used.

This concludes a brief overview of the CSMRPAE theory --- we may use Eqs.~\ref{eq:MRPA_gauge_whole_expression},~\ref{eq:eliashberg_Z},~\ref{eq:eliashberg_phi} to compute the Eliashberg coupling constants. Results are shown in Fig.~\ref{fig:lambdacouplings} of the main text and in Section \ref{sec:CECEresults_numeric}.

\subsection{Why not go to MMRPA?}
The MMRPA is an additional correction above the MRPA \cite{simon_MRPA}. In MMRPA, magnetic flux is attached to the composite fermions, recovering the physics of the particles undertaking cyclotron motion and therefore carrying orbital magnetization, which is lost in the mean-field approximation. We choose to not apply that correction here: the MMRPA is designed to study the response to the \textit{external} gauge field, $A_\mu$, but we are really interested in the response to fluctuations of the \textit{statistical} gauge field, $a_\mu$. Microscopic insight into how cyclotron motion gives rise to additional currents and changes the effective external field may be used to develop the MMRPA as a response to the external gauge field, but such arguments do not necessarily apply to the statisitcal gauge field. For this reason, we believe the MMRPA to not be an appropriate correction in this computation.   A second way to justify our choice is to state that we are pairing fermions that are undressed of their orbital angular momentum. 
The MMRPA and MRPA differ only in the way they couple to external transverse gauge fields.   However, the coupling between layers that drives the pairing can only be from the conventional (longitudinal) Coulomb interaction.  Thus, the MRPA should fully capture the pairing glue.   Readers may find it useful to consider the discussion of the MRPA and MMRPA given in the Appendix of Ref.~\cite{LevinSon}.   Here the MRPA is viewed as a coupling to an auxiliary velocity field, and the MMRPA again only modifies the coupling to the external transverse gauge field.

\section{Different Divergences in CE-CE pairing versus CE-CH pairing when evaluated in RPA or nonzero $m_b$ MRPA}
\label{sec:CECHdivergences}

In this section we look at the RPA calculations of Isobe-Fu\cite{Fu} and  R\"uegg-Chaudhary-Slager\cite{Slager} and show that, while the divergent terms for the quasiparticle residue coupling constants $\lambda_Z$ are equal between the two cases  (CE-CE vs CE-CH), the divergent terms in the pairing coupling constants $\lambda_\phi$ differ. Thus any comparison between the strength of the two pairing instabilities will depend on the cutoff prescription. We also show that the divergent terms are still present in the MRPA if $m_b>0$, and disappear as a power of $m_b$, meaning there are no cutoff-based issues in the $m_b=0$ MRPA. We formally work in the MRPA, noting that the RPA may be recovered exactly by setting $m_b=m^*$.

Isobe and Fu \cite{Fu} observe that the Eliashberg couplings $\lambda_Z$ and $\lambda_\phi$ diverge as a function of the low-momentum cutoff, $q_c$, with $\lambda_Z\propto1/q_c$ and $\lambda_\phi\propto\log q_c$. We show that these divergences are multiplied by a power of $m_b$, meaning they go away in the limit $m_b\to 0$. In this section we focus on the low-momentum limit, $q\ll k_F$, but allow $\omega_m$ to take any value. We choose units with $k_F=1$ and $e=1$. In this limit \cite{Fu}, we have
\begin{align}
    \Pi_{00}(q,i\omega_m)&\approx-\frac{\epsilon_F}{2\pi}\frac{q^2}{\omega_m^2}\left(1+\mathcal{O}(q^2/\omega_m^2)\right) \label{eq:SmallQPI00}\\
    \Pi_{11}(q,i\omega_m)&\approx\frac{\epsilon_F}{2\pi}\left(1+\mathcal{O}(q^2/\omega_m^2)\right) \label{eq:SmalQPI11}
\end{align}
From Eq.~\ref{eq:K_RPA_DEF}--\ref{eq:MRPA_F1_def}, we get
\begin{equation}
    \left(K^\text{MRPA}_\pm(q,i\omega_m)\right)^{-1}_{\mu\nu}=\begin{pmatrix}
        -\frac{2\pi}{\epsilon_F}\frac{\omega_m^2}{q^2} & 0 \\
        0 & \frac{2\pi}{\epsilon_F}
    \end{pmatrix}\left(1+\mathcal{O}(q^2)\right)
    -\begin{pmatrix}
        V_\pm(q) & \frac{4\pi}{q} \\
        \frac{4\pi}{q} & 0
    \end{pmatrix}
    +
    \frac{m^*-m_b}{n_e}
    \begin{pmatrix}
        \frac{\omega_m^2}{q^2} & 0 \\
        0 & -1
    \end{pmatrix}
\end{equation}
where $V_\pm(q)=V_{11}(q)\pm V_{12}(q)$. Using that $k_F=1$, we have $n_e=1/4\pi$ and $\epsilon_F=1/2m^*$. This implies
\begin{equation}
    K^\text{MRPA}_{\pm,\mu\nu}(q,i\omega_m)=\frac{-1}{\frac{(4\pi)^2}{q^2}+4\pi \Tilde{m}_b\left[V_\pm(q)+4\pi \Tilde{m}_b\frac{\omega_m^2}{q^2} \right]}
    \begin{pmatrix}
        4\pi \Tilde{m}_b & \frac{4\pi}{q} \\
        \frac{4\pi}{q} & -4\pi \Tilde{m}_b \frac{\omega_m^2}{q^2}- V_\pm(q)
    \end{pmatrix},
\end{equation}
where $\Tilde{m}_b=m_b+\mathcal{O}(m^*q^2)$ is defined to absorb the inaccuracy in our approximation of $\Pi$.

Using Eq.~\ref{eq:D_rpa_from_K_rpa}, we get 
\begin{equation} \label{eq:good_low_momentum_gauge_propagator}
    \mathcal{D}^\text{MRPA}_{\pm,\mu\nu}(q,i\omega_m)=\frac{1}{(4\pi)^2+4\pi \Tilde{m}_b\left[q^2V_\pm(q)+4\pi \Tilde{m}_b \omega_m^2\right]}\begin{pmatrix}
        V_\pm(q)(4\pi \Tilde{m}_b \omega_m)^2 + 4\pi \Tilde{m}_b \omega_m^2 \frac{(4\pi)^2}{q^2} & \frac{4\pi}{q}(4\pi \Tilde{m}_b\omega_m)^2 \\
        \frac{4\pi}{q}(4\pi \Tilde{m}_b\omega_m)^2 & 0.
    \end{pmatrix}
    +\mathcal{O}(1).
\end{equation}
When computing $\lambda_Z$ from Eq.~\ref{eq:eliashberg_Z} we are concerned with terms of the type $\mathcal{D}^\text{MRPA}_{+,\mu\nu}+\mathcal{D}^\text{MRPA}_{-,\mu\nu}$, which can be approximated to lowest two orders in $q$ as
\begin{equation} \label{eq:z_integrand}
    \begin{split}
        &\mathcal{D}^\text{MRPA}_{+,\mu\nu}(q,i\omega_m)+\mathcal{D}^\text{MRPA}_{-,\mu\nu}(q,i\omega_m)=\\
        &=\frac{2}{(4\pi)^2\left(1+\Tilde{m}_b^2\omega_m^2\right)}\begin{pmatrix}
        V_{11}(q)(4\pi m_b \omega_m)^2 + 4\pi m_b \omega_m^2 \frac{(4\pi)^2}{q^2} & \frac{4\pi}{q}(4\pi m_b\omega_m)^2 \\
        \frac{4\pi}{q}(4\pi m_b\omega_m)^2 & 0
    \end{pmatrix}
    +\mathcal{O}(1)    
    \end{split}
\end{equation}
Since $\Tilde{m}_b=m_b+\mathcal{O}(m^*q^2)$, the most divergent term is $\Tilde{m}_b/q^2=m_b/q^2+\mathcal{O}(1)$, and we may simply replace all the $\Tilde{m}_b$ by $m_b$. The leading order divergence in $\mathcal{D}_{00}$ is discussed in Section VI-C of \cite{HLR}.   See section \ref{sub:analysisofdivergence} below for a more detailed discussion of this divergence and the comparison with the discussion of Halperin-Lee-Read\cite{HLR}. 
Equation~\ref{eq:eliashberg_Z} shows that for momentum cutoff $q_c$,
\begin{equation}
    \lambda_Z\approx 
    A\frac{m_b\omega_m^2}{1+m_b^2\omega_m^2}\frac{1}{q_c} + B \frac{m_b^2\omega_m^2}{1+m_b^2\omega_m^2}\log q_c + \mathcal{O}(q_c^0), \label{eq:low_q_lambda_z}
\end{equation}
with $A, B$ being $\mathcal{O}(1)$ constants. Note that in these units, $\omega_c=1/m_b$. 
At a given frequency, when $m_b\to0$, there is no divergence in $\lambda_Z$. At finite $m_b$, the leading divergence is $\sim 1/q_c$, with the strength proportional to $m_b$ when $\omega_m\ll\omega_c=1/m_b$, but proportional to $1/m_b$ when $\omega_m\gg\omega_c=1/m_b$.

To compute $\lambda_\phi^{(l)}$ via Eq.~\ref{eq:eliashberg_phi}, we need to evaluate terms of the type $\mathcal{D}^\text{MRPA}_{+,\mu\nu}-\mathcal{D}^\text{MRPA}_{-,\mu\nu}$. Again approximating up to zeroth order in $q$, we find (since $V_{12}\sim 1/q$ at small $q$)
\begin{equation} \label{eq:phi_integrand}
    \begin{split}
        &\mathcal{D}^\text{MRPA}_{+,\mu\nu}(q,i\omega_m)-\mathcal{D}^\text{MRPA}_{-,\mu\nu}(q,i\omega_m)=\\
        &=\frac{2\left(\Tilde{m}_b\omega_m\right)^4 }{\left[1+\Tilde{m}_b^2\omega_m^2\right]^2}\begin{pmatrix}
        V_{12}(q) & 0 \\
        0 & 0
    \end{pmatrix}
    +\mathcal{O}(1)    
    \end{split}
\end{equation}
The most divergent term is $\sim 1/q$, so replacing $\Tilde{m}_b$ with $m_b$ only introduces an error $\mathcal{O}(q)$. With a momentum cutoff $q_c$, we get from Eq.~\ref{eq:eliashberg_phi} 
\begin{equation} \label{eq:low_q_lambda_phi}
    \lambda_\phi^{(l)}\approx c_1 C \frac{m_b^4\omega_m^4}{\left(1+m_b^2\omega_m^2\right)^2} \log q_c + \mathcal{O}(1),
\end{equation}
with $C$ a $\mathcal{O}(1)$ constant. So the leading logarithmic divergence goes away as $m_b\to0$. At finite $m_b$, the strength of the divergence is $\sim m_b^4\omega_m^4=(\omega_m/\omega_c)^4$ when $\omega_m m_b=\omega_m/\omega_c\ll1$: in the opposite limit $\omega_m m_b\gg1$, we have $\lambda_\phi^{(l)}\approx c_1C\log q_c+\mathcal{O}(1)$. The important part of the result Eq.~\ref{eq:low_q_lambda_phi} is that the divergent term is independent of $l$, but is dependent on $c_1$, i.e., on the type of coupling: it is positive for CE-CE, but negative for CE-CH. This confirms, as suggested by Isobe-Fu \cite{Fu} that for CE-CE coupling, different-$l$ channels may be compared by introducing a cutoff, as the divergent terms are equal anyway. The same holds for different $l$-channels of CE-CH, but does not hold for comparing CE-CE to CE-CH. In the case of comparing CE-CE to CE-CH, the divergent terms are different (in fact exactly opposite) so the precise values of the couplings may not be compared directly in a meaningful way. This analytically supports the numerical results shown in Fig.~\ref{fig:slager_cutoff_dependance} of Section \ref{sec:slagerCutoff}, where the ordering in strength of the coupling channels and the difference between their strengths depends on the cutoff choice. The only case where a CE-CE to CE-CH comparison is meaningful is when $m_b\to0$ where the divergence disappears.

\section{Detailed analysis of divergences, even at very high $\omega_m$}

\label{sub:analysisofdivergence}

We have successfully eliminated divergences that stem from the cyclotron mode, by pushing the cyclotron mode off to infinity in the $m_b \rightarrow 0$ limit.   One might worry whether, no matter how large the cyclotron scale is, divergences at  (or above) that scale may be problematic.   This potential concern encouraged us to look more closely at these divergences.  

First, we find that the divergences  obtained by the approaches of  Refs.~\cite{Bonesteel,Fu,Slager}, (i.e., all cases of divergences by prior groups mentioned in the main text) are greatly exaggerated by details of the approximation method used.  A more careful evaluation of the Eliashberg equations shows that for nonzero value of $m_b$ the divergence of the nonanomalous self-energy stemming from the cyclotron mode is actually only $\log q_c$ not $1/q_c$.    Furthermore, this log divergence is precisely the ultraviolet divergence discussed in detail by Halperin-Lee-Read\cite{HLR} (section VI.C) which stems from the vanishing overlap between wavefunctions with and without the singular gauge transformation.   This type of divergence can be fairly safely ignored, as it has been in the past\cite{HLR}.    A similar analysis shows that the anomalous part of the self-energy is actually not divergent at all.   With this revised understanding of the divergences in mind, we expect that pushing the cyclotron frequency up to infinite energy should leave us with a well behaved low energy theory.   

Roughly, the error by prior works\cite{Fu,Slager,Bonesteel} in evaluation of the infrared divergences stems from the restriction of a momentum integral in the self-energy to lie precisely on the Fermi surface.   For the non-anomalous self-energy this gives a divergent one-dimensional integral $\int_{q_c} dq \, V_{ex}(q) \sim 1/q_c $ where $V_{ex} \sim 1/q^2$ is the most singular part of the gauge interaction.  However, the integral is not really one-dimensional, but rather is an integral over a strip near the Fermi surface whose width is set by a (potentially small but) finite energy scale related to the frequency scale of the pairing attraction.   At small enough $q$ this integral should always  be treated as two-dimensional yielding instead  $\int_{q_c} d^2q \, V_{ex}(q) \sim \log q_c$, a much less serious divergence, and as mentioned above, one that we understand well\cite{HLR}.  Using a similar analysis, the divergence in the anomalous self-energy was previously described as a one dimensional integral  $\int_{q_c} dq \, V_c(q) \sim \log q_c$ where $V_c \sim 1/q$.  Again the divergence comes from treating the integral as strictly one-dimensional, whereas a more proper two-dimensional treatment finds the integral $\int_{q_c} d^2 q \, V_c(q)$ to be non-divergent. 

Let us now take a more detailed look at some of these divergences for nonzero $m_b$.   Before so doing, we should recall the derivation of the coupling constants Eqs.~\ref{eq:lambdaz}-\ref{eq:lambdaphi}.  Note in particular that we have used inequality Eq.~\ref{eq:inequality} to restrict the integration to the Fermi surface.   However, the left hand side of this inequality is never arbitrarily small, so the resulting delta-function obtained
in Eqs.~\ref{eq:lambdaz}-\ref{eq:lambdaphi} is never infinitely sharp.  In fact, we should think of the delta function $\delta(\xi_{\bf q+k})$ as having a thickness of 
$$
  \delta \xi_{n,m} =  \sqrt{Z^2_{n+m} (\epsilon_n + \omega_m)^2  + |\phi_{n+m}^{(l)}|^2} 
$$
or equivalently the distance perpendicular from the Fermi surface is smeared by
$$
 \delta Q_{\perp,n,m} =  \frac{1}{v_F} \sqrt{Z^2_{n+m} (\epsilon_n + \omega_m)^2  + |\phi_{n+m}^{(l)}|^2} 
$$
with $v_F$ the Fermi velocity.  For $q$ greater than this scale, we should think of the $q$ integral in Eqs.~\ref{eq:lambdaz}-\ref{eq:lambdaphi} as being one-dimensional as we have done in Eqs.~\ref{eq:eliashberg_Z}-\ref{eq:eliashberg_phi}.  However, for $q$ smaller than this scale the $q$ integral should be thought of as being two-dimensional.   For each value of $n$ and $m$ in Eqs.~\ref{eq:Eliashberg01}-\ref{eq:Eliashberg02} this cutoff scale is different, however, it is always nonzero for any finite $T$ or $\omega_m$.

This analysis suggests it should be possible to pursue Eliashberg Chern-Simons theory without serious divergences even for nonzero values of $m_b$, although this would require a more careful analysis of the relevant two-dimensional integrals, which is beyond the scope of this work. Instead we focus in this work on the $m_b \rightarrow 0$ limit such that all problematic divergences are pushed away to infinite frequencies and we can safely work with the one-dimensional Fermi surface integrals. 

\section{Comments on Chern-Simons theory and Landau level projection}

\label{sec:LLprojection}

The Chern-Simons fermion (CE) approach to fractional quantum Hall effects\cite{LopezFradkin,HLR} begins by making a singular gauge transformation that then represents each electron as a fermion bound to an even number of (Chern-Simons) flux quanta.   This transformation (sometimes called a Chern-Simons transformation) 
is exact.    Once one makes the exact transformation, one is typically forced to make a mean field approximation of some sort, and then attempt to  include the effects of fluctuations around mean field using approaches such as RPA.   It is at the mean-field step that the calculation is no longer exact.   Nonetheless, we may hope that an approach to including fluctuations, such as MRPA\cite{SimonHalperinMRPA,simon_MRPA} may still describe the physics, at least roughly, for any value of $m_b$ (which we treat as a separate parameter from $m^*$ which is set  by the interaction scale).  In particular, we may hope that in the limit of $m_b$ going to zero  we would at least roughly represent the physics restricted to the lowest Landau level (this has been tested with good results, for example, in Ref.~\cite{HeSimonHalperin}), and for finite $m_b$ we would expect the MRPA to incorportate the effects of Landau level mixing. 

The idea of attaching Chern-Simons flux to {\it holes} within the lowest Landau level (CH) was suggested, and derived,  by Ref.~\cite{AntiCFLiquid}.    While the resulting field theory seems valid as a low energy long wavelength description, it is not on the same footing as the above CE Chern-Simons approach, which is in principle an exact description before the mean-field approximation is made.    The CH model is not exact in the same way since one cannot even define a hole degree of freedom except within the finite dimensional Hilbert space of a single Landau level.  Further, once we project to a single Landau level, if we try to make the exact singular gauge transformation, the system would no longer remain within that Landau level.    (We note in passing that for a fractional Chern insulator with finite dimensional Hilbert space, one could in principle attach flux to holes in a well-defined way without any projection to a single band.  However, this would not result in the correct mean field state that we want for the CH Fermi liquid.)   

The CH model is strictly equivalent to a physical system of charge $+e$ fermions (compared to charge $-e$ electrons) in a magnetic field.   When projected to a single Landau level, the CH model is equivalent to a theory of holes in the single Landau level of electrons (which is why we use this model).  However, when Landau level mixing is included, the system of charge $+e$ fermions and the system of charge $-e$ electrons are inequivalent.

As mentioned above, in the limit that $m_b$ is taken to zero, and one can think of the system of being projected to a single Landau level and the concept of particle-hole conjugation is well defined within the Hilbert space of the single Landau level.   In this case, the CH MRPA theory also becomes an equally good description of the system as the CE MRPA.

\section{Favoured channels for CE-CE and CE-CH pairing in MRPA Chern-Simons Eliashberg Theory}
\label{sec:CECEresults_numeric}

\begin{figure}[h]
    \centering
    \begin{subfigure}{0.48\textwidth}
      \centering
      \includegraphics[width=2.5in]{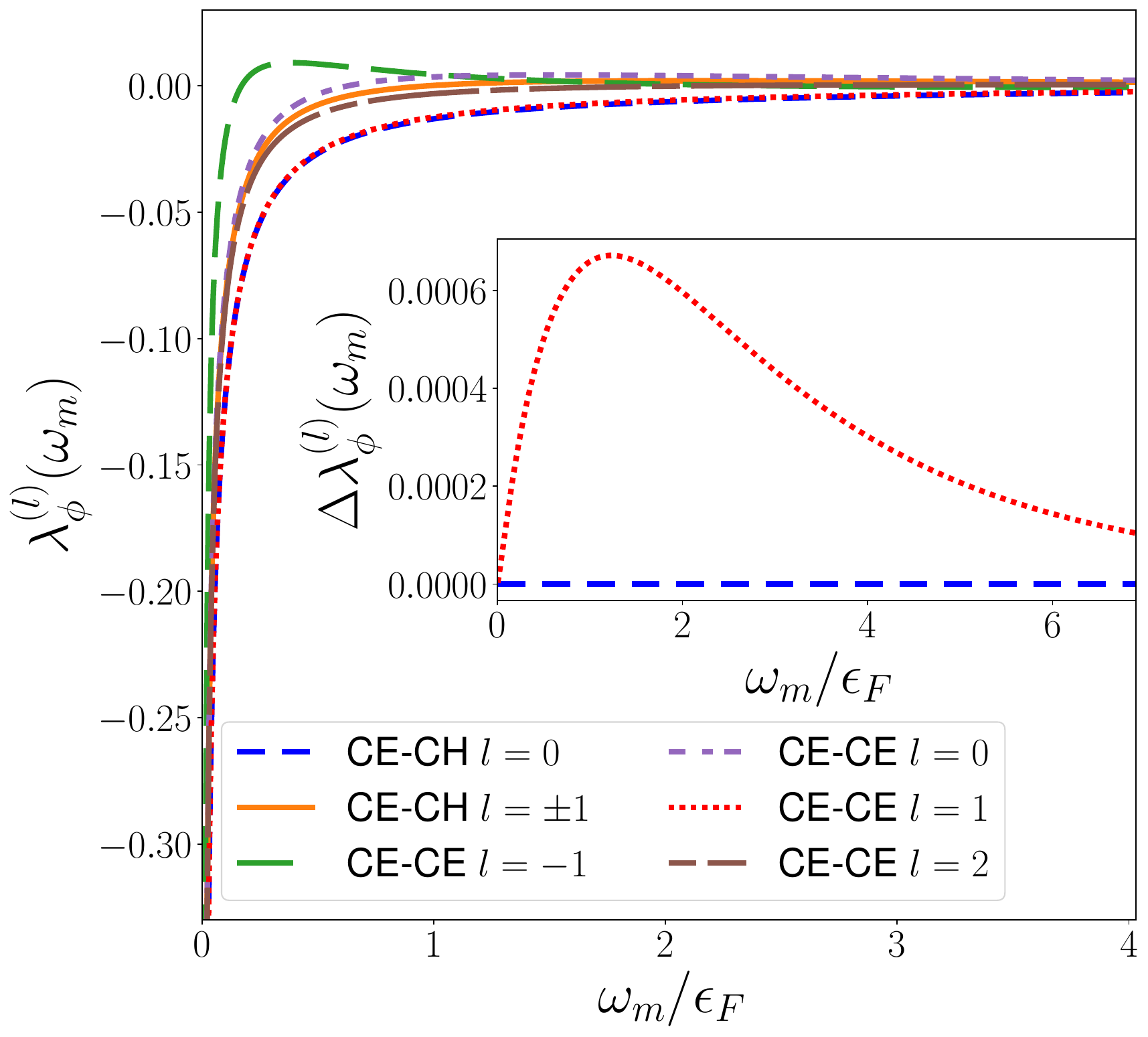}
      \caption{$d/l_B=1$}
      \label{fig:MRPA_d_1_supplement}
    \end{subfigure}
    \begin{subfigure}{0.48\textwidth}
      \centering
      \includegraphics[width=2.5in]{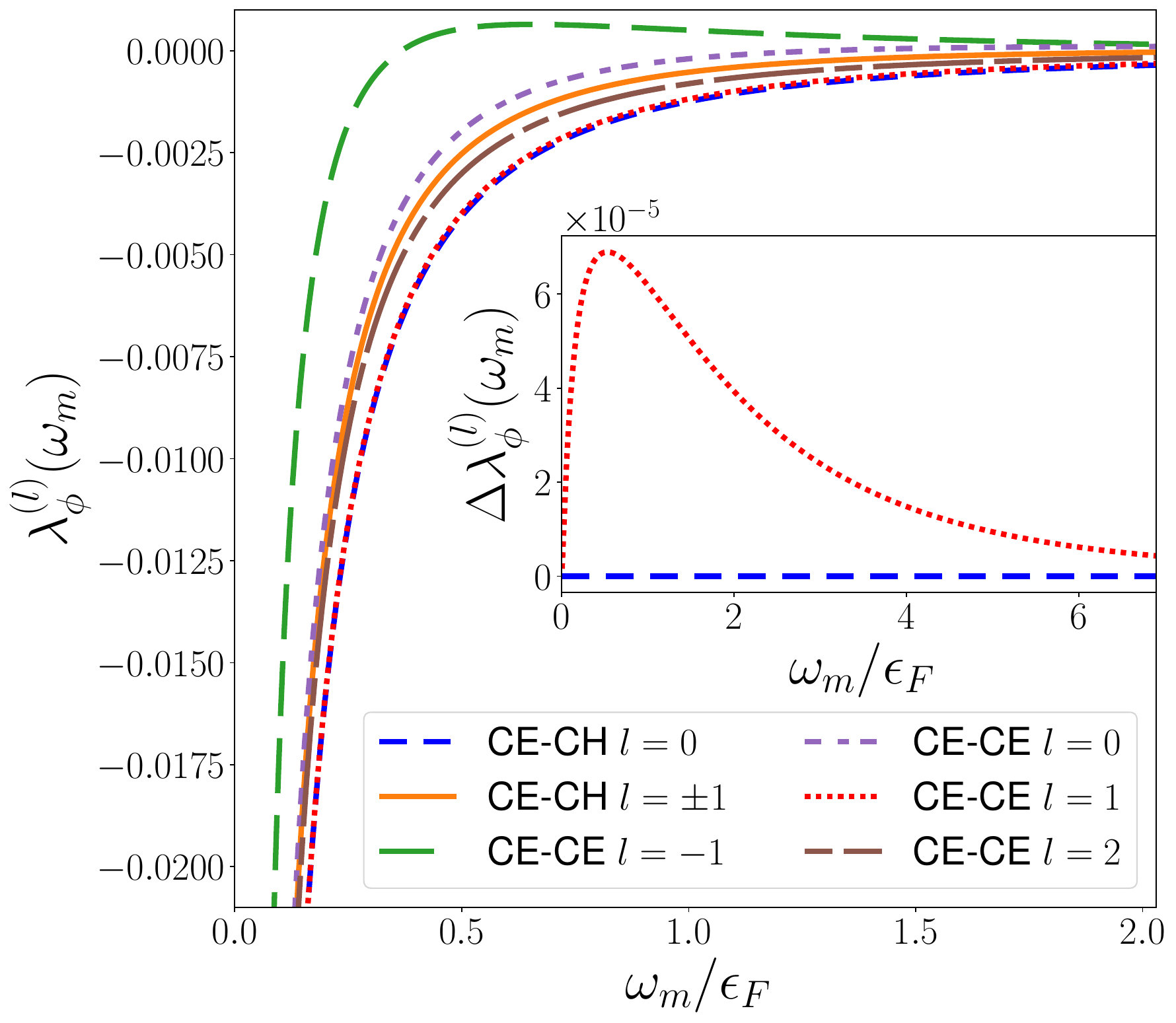}
      \caption{$d/l_B=3$}
      \label{fig:MRPA_d_3_supplement}
    \end{subfigure}
    \caption{The results for $\lambda_\phi^{(l)}(\omega_m)$ within the $m_b=0$ MRPA, showing unambiguous results for a large range of $\omega_m$ for two different values of $d/l_B$. In CE-CE coupling, $l=1$ is always favoured, and in CE-CH coupling, $l=0$ is favoured. The two are very similar, but not exactly equal in strength. The insets show the difference between them, magnified.}
    \label{fig:MRPA_big_range}
\end{figure}

We evaluate the pairing Eliashberg constants $\lambda_{\phi}^{(l)}(\omega_m)$ by numerically integrating Eq.~\ref{eq:eliashberg_phi} within the MRPA with $m_b=0$, using Eq.~\ref{eq:MRPA_gauge_whole_expression} for the gauge propagator $\mathcal{D}_{\mu\nu}$. Note that as there is no divergence now, there is no need to introduce a cutoff. The results in Fig.~\ref{fig:MRPA_big_range} show that for CE-CE coupling, $l=1$ is now favoured unambigously, with $d=l_B$ and $d=3l_B$ shown. Even if $d$ is made larger, or the frequency range extended, $l=1$ is always the strongest coupling, unlike in the RPA results presented in Section \ref{sec:Isobe}. Another important thing to note is that $\lambda_{\phi}^{(l)}(\omega_m)\to0$ always for $\omega\gg\epsilon_F$. With RPA we would have $\lambda_{\phi}^{(l)}(\omega_m)>0$ for CE-CE and $\lambda_{\phi}^{(l)}(\omega_m)<0$ for CE-CH at high frequency, as discussed in section \ref{sec:CECHresults_analytic}.

As evident from Eq.~\ref{eq:eliashberg_phi}, $l$ and $-l$ are degenerate for CE-CH coupling. For CE-CE coupling $l=0$ and $l=2$ are similar in strength, both similar to $l=1$ in CE-CH, and in CE-CE, $l=-1$ is similar in strength to $l=3$, both similar to CE-CH $l=2$. If the coupling were exactly symmetric between CE-CE and CE-CH, we would expect $(l+1)$-wave of CE-CE to coincide with $l$-wave of CE-CH, coinciding with the $-l$-wave CE-CH and $(-l+1)$-wave CE-CE too. We see this degeneracy only approximately. If the adjustment of Section \ref{sub:adjustment} is made to the MRPA, these quadruplets are exactly degenerate in strength.

By far the most precise degeneracy is seen between $l=0$ CE-CH and $l=1$ CE-CE, as shown in Fig.~\ref{fig:MRPA_big_range}: the coupling strengths follow each other closely, with the CE-CH $l=0$ being very slightly stronger. The difference seen here is over 2 orders of magnitude smaller than the difference seen within the RPA in \cite{Slager}.   The difference is sufficiently small that we do not believe it is physically meaningful, given the rough level of this calculation. 

\subsection{Different interaction potentials}
\label{sub:potentials}

\begin{figure}[h]
    \centering
    \begin{subfigure}{0.48\textwidth}
      \centering
      \includegraphics[width=2.5in]{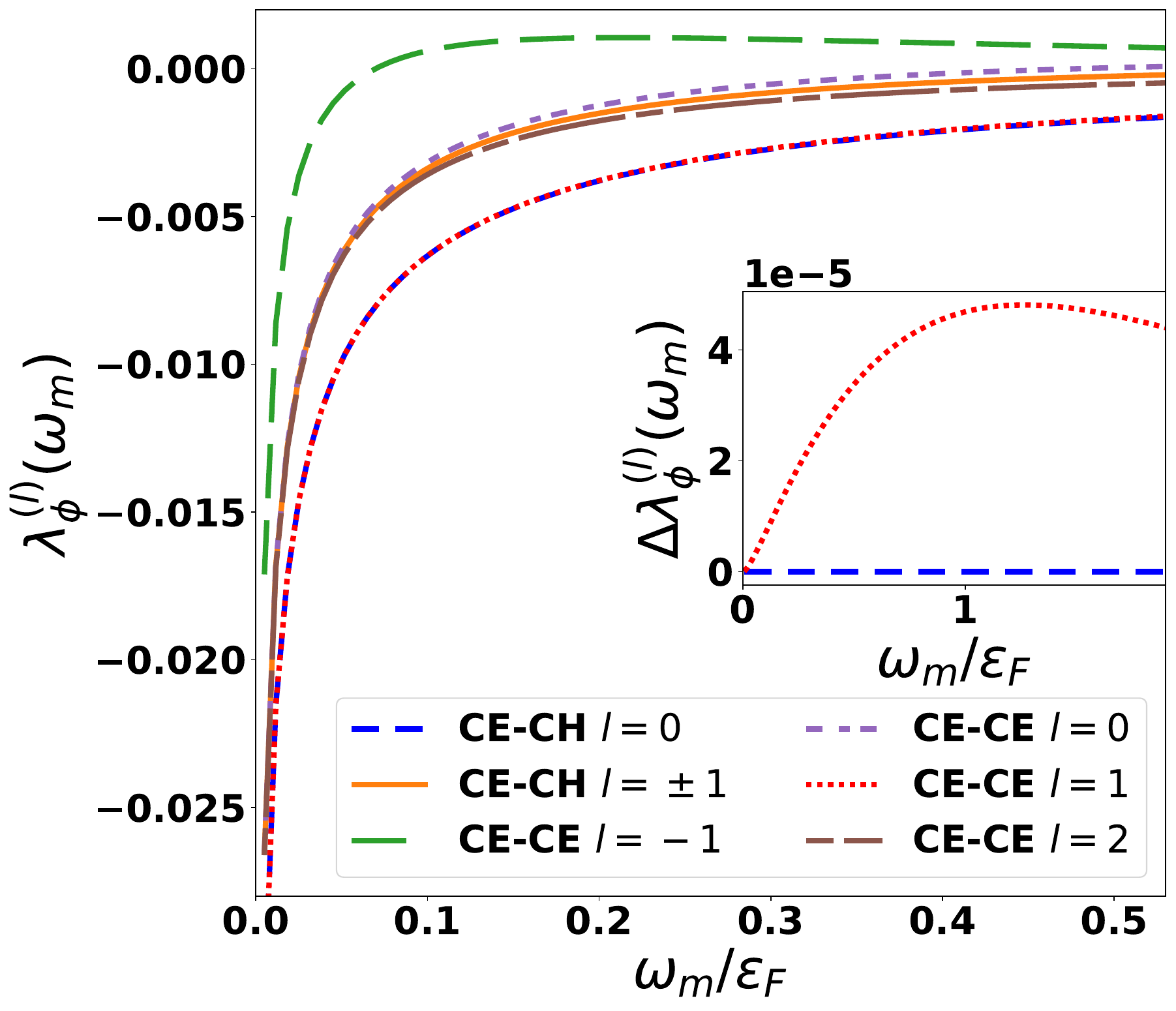}
      \caption{$A_{12}=0.2\epsilon_F$}
      \label{fig:gauss_weak_12}
    \end{subfigure}
    \begin{subfigure}{0.48\textwidth}
      \centering
      \includegraphics[width=2.5in]{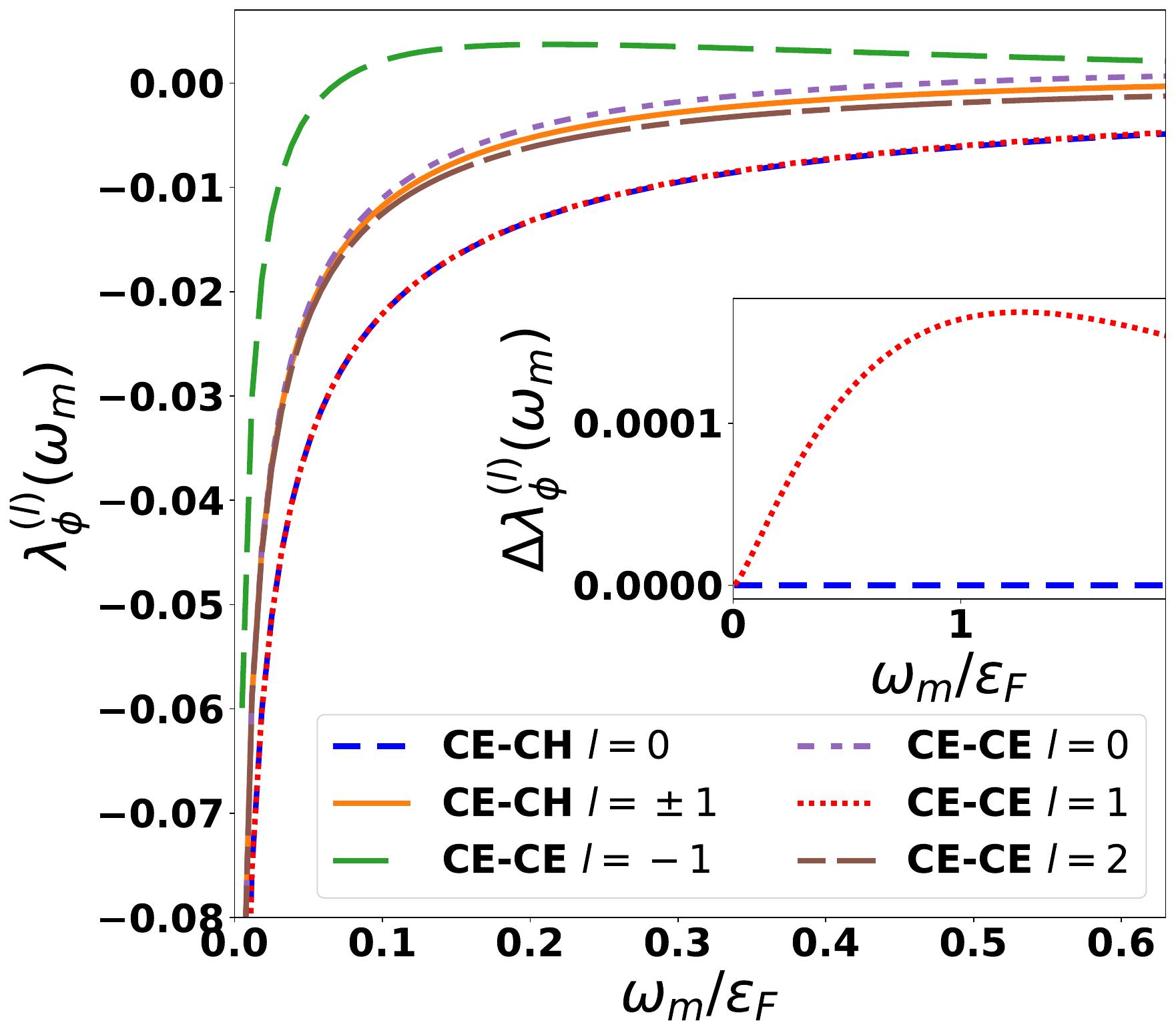}
      \caption{$A_{12}=0.7\epsilon_F$}
      \label{fig:gauss_strong_12}
    \end{subfigure}
    \caption{Results for $\lambda_\phi^{(l)}(\omega_m)$ for a Gaussian interaction of width $w=l_B$ with $A_{11}=\epsilon_F$ and $A_{12}\in\{0.2\epsilon_F,0.7\epsilon_F\}$. The insets show the difference between the two favoured channels, magnified.}
    \label{fig:gaussian_potential_lambda_phi}
\end{figure}

We also investigate what happens if an external potential different to the Coulomb $V(q)\sim q^{-1}$ is used. With a Gaussian potential, $V(q)=Ae^{-\frac{1}{2}(qw)^2}$, with $w$ controlling the width and $A$ the strength (both $V_{11}$ and $V_{12}$ are taken to have this form, but possibly different strength and width), we see qualitatively the same results as for the Coulomb potential, that is $l=1$ favoured for CE-CE and $l=0$ for CE-CH, with the two being very close to degenerate. We have tried $w/l_B\in(0.5,5)$ for both $V_{11}$ and $V_{12}$, and $A/\epsilon_F\in(1,10)$ for $V_{11}$ and $A/\epsilon_F\in(e^{-5},10e^{-1})$ for $V_{12}$. For completeness, the coupling constants for $A_{11}=\epsilon_F$, $A_{12}\in\{0.2\epsilon_F,0.7\epsilon_F\}$ and $w=l_B$ are shown in Fig~\ref{fig:gaussian_potential_lambda_phi}.

Further, for a screened potential $V(q)\sim (q+q_0)^{-1}$ (with $q_0\leq0.5k_F$), longer range potentials $V(q)\sim q^{-2+\eta}$ (so $V(r)\sim r^{-\eta}$) for $\eta\in (0,1)$, potentials projected into any of the first three Landau Levels (by replacing $V(q)\to V(q)\left[L_n\left(\frac{1}{2}l_B^2q^2\right)\right]^2$ \cite{Duncan1987}), or cases where finite sample width is accounted for by multiplying potentials by a Gaussian factor $e^{-\frac{1}{2}q^2w^2}$ (for $w\leq4l_B$), we similarly see good qualitative agreement with the above results in the Coulomb case.

In short: numerical integration of Eq.~\ref{eq:eliashberg_phi} shows that $l=1$ is unambiguously favoured for CE-CE coupling, and confirms that $l=0$ is favoured for CE-CH coupling. It also shows that the two are extremely similar, but not precisely equal in strength. This holds for a variety of rotationally symmetric potentials.

\subsection{Other filling fractions}
\label{sub:otherfilling}
\begin{figure}[h]
    \centering
    \begin{subfigure}[b]{0.4\textwidth}
        \centering
        \includegraphics[width=2.5in]{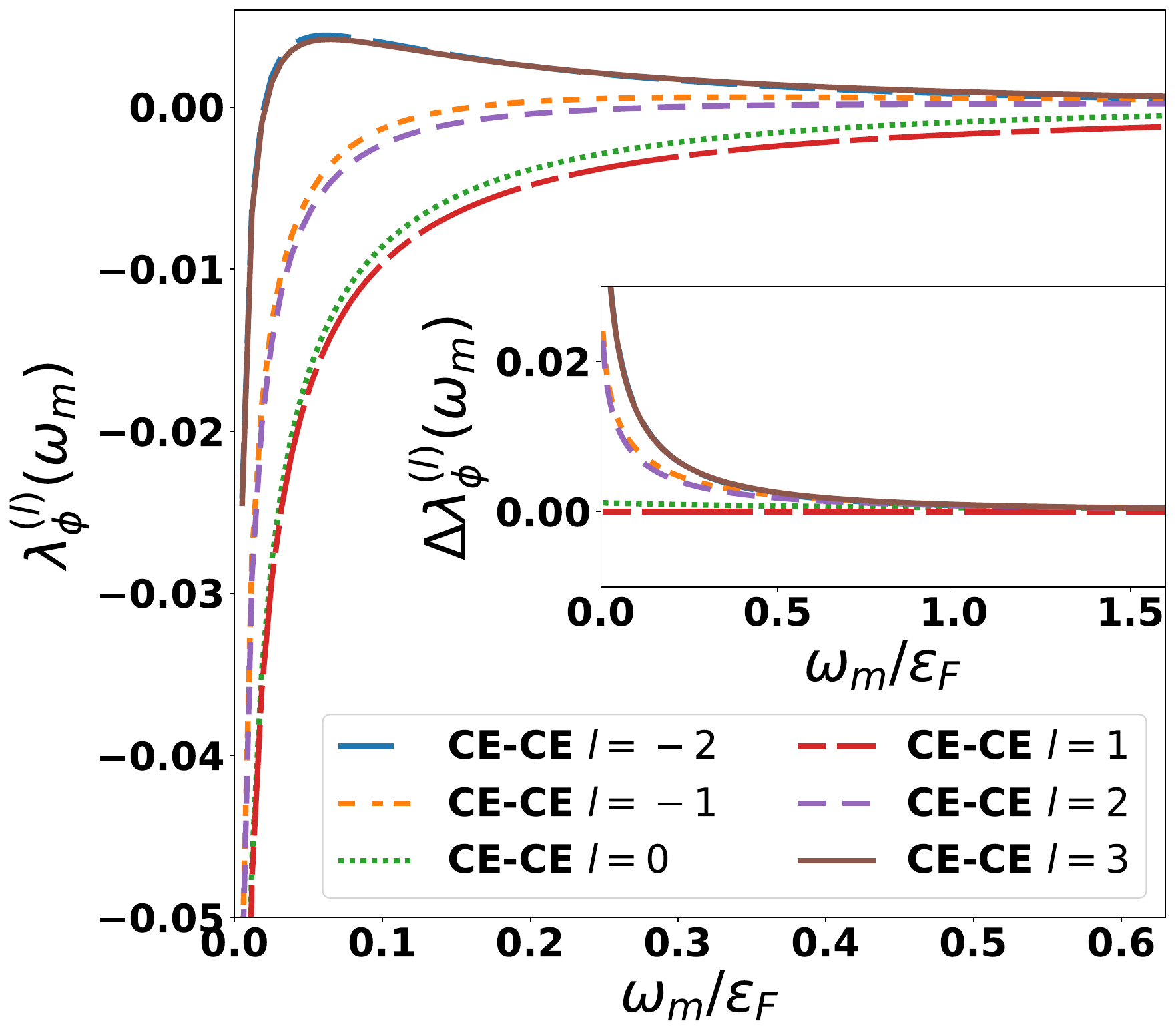}
        \caption{$\nu=\frac{1}{4}+\frac{1}{4}$ as CE-CE coupling for $\Tilde{\phi}=4$}    
        \label{fig:cece_4}
    \end{subfigure}
    \hfill
    \begin{subfigure}[b]{0.4\textwidth}  
        \centering 
        \includegraphics[width=2.5in]{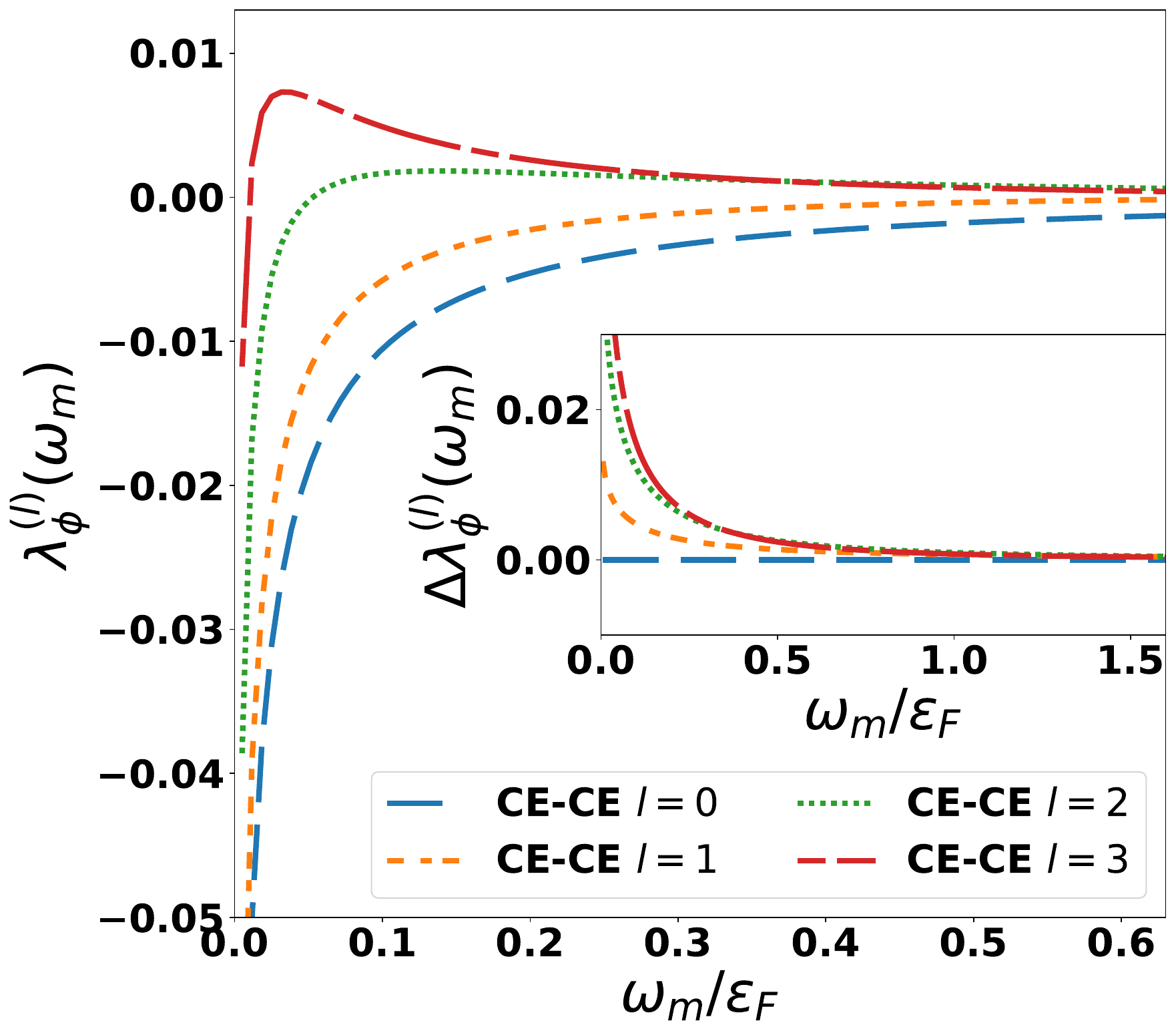}
        \caption{$\nu=\frac{1}{4}+\frac{3}{4}$ as CE-CH coupling for $\Tilde{\phi}=4$}    
        \label{fig:cech_4}
    \end{subfigure}
    \vskip\baselineskip
    \begin{subfigure}[b]{0.4\textwidth}   
        \centering 
        \includegraphics[width=2.5in]{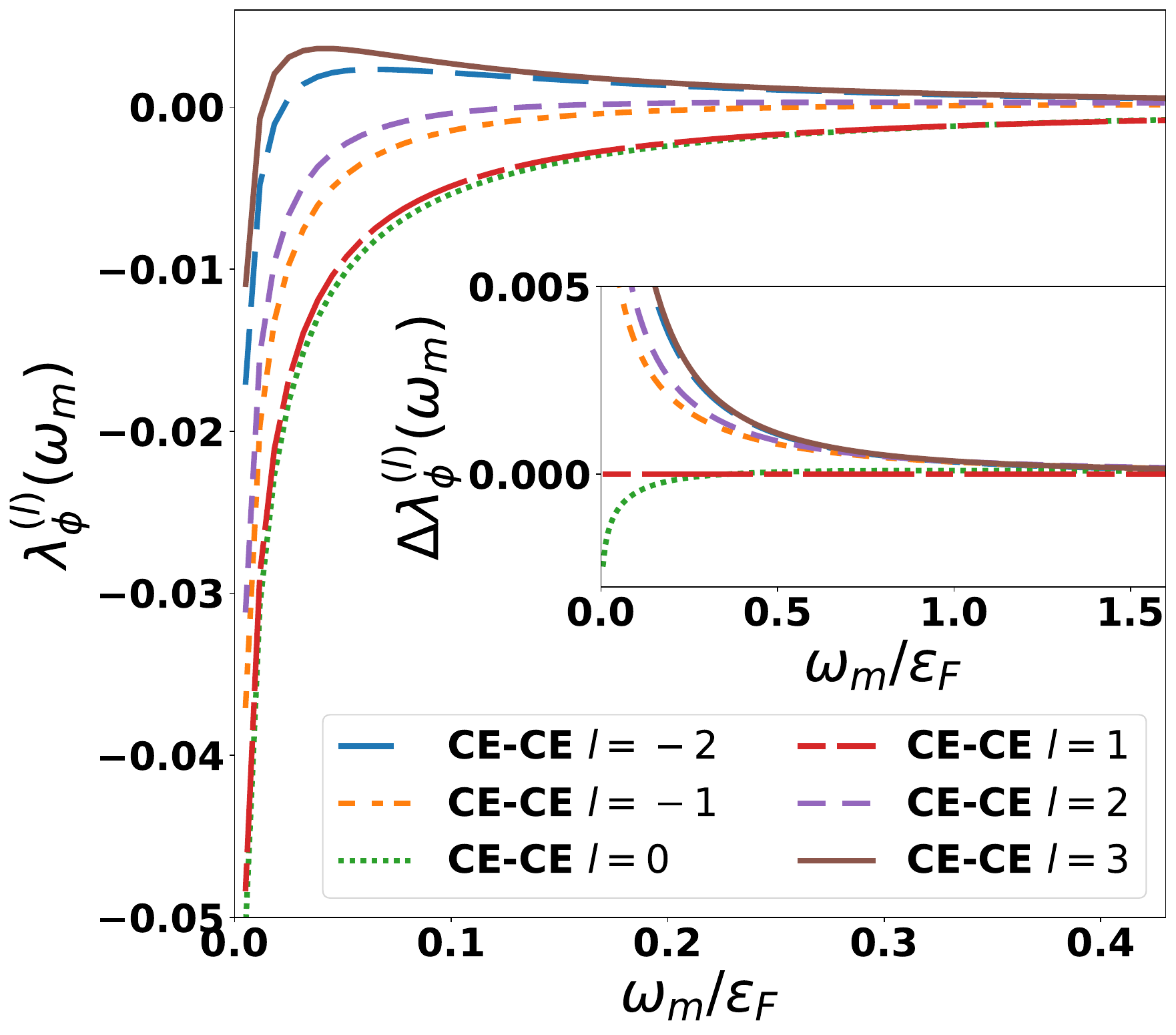}
        \caption{$\nu=\frac{1}{6}+\frac{1}{6}$ as CE-CE coupling for $\Tilde{\phi}=6$}    
        \label{fig:cece_6}
    \end{subfigure}
    \hfill
    \begin{subfigure}[b]{0.4\textwidth}   
        \centering 
        \includegraphics[width=2.5in]{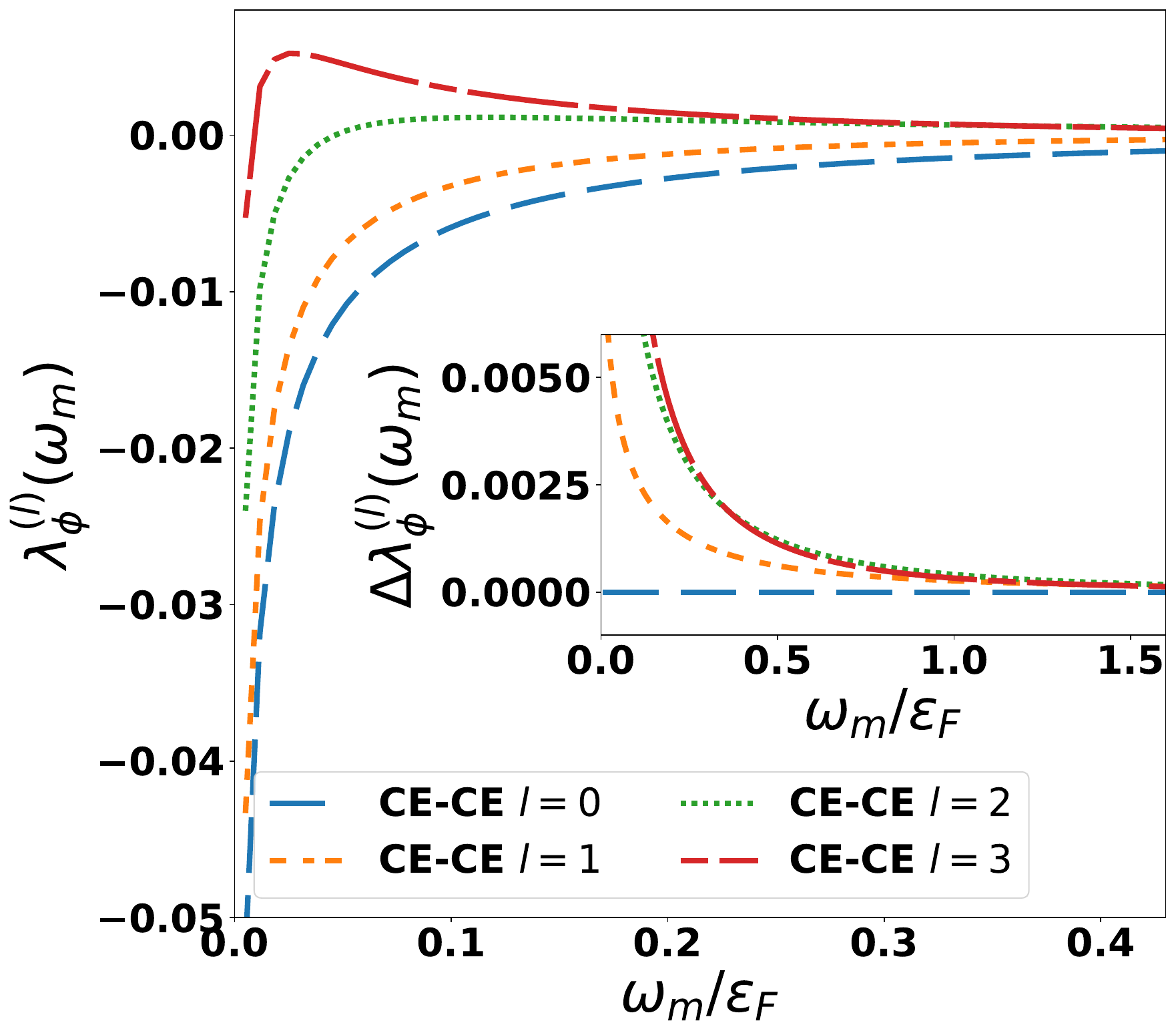}
        \caption{$\nu=\frac{1}{6}+\frac{5}{6}$ as CE-CH coupling for $\Tilde{\phi}=6$}    
        \label{fig:cech_6}
    \end{subfigure}
    \caption{The results for $\lambda_\phi^{(l)}(\omega_m)$ within the $m_b=0$ MRPA for several other compressible filling fractions. } 
    \label{fig:other_filling_fractions}
\end{figure}

Our method allows for an easy generalisation to compressible QHE states away from $\nu=\frac{1}{2}+\frac{1}{2}$. Re-write the pre-factor of the Chern-Simons term in Eq.~\ref{eq:euclidian_action} as $\frac{1}{2\pi\Tilde{\phi}}$, where $\Tilde{\phi}$ is to be understood as the number of flux quanta attached to each electron \cite{Fu}. Above, we were considering the pre-factor $\frac{1}{4\pi}$, which corresponds to standard HLR theory with $\Tilde{\phi}=2$. Considering also other even integers for $\Tilde{\phi}$ allows us to access states of $\nu=\frac{1}{\Tilde{\phi}}+\frac{1}{\Tilde{\phi}}$ with CE-CE pairing and $\nu = \frac{1}{\Tilde{\phi}}+\frac{\Tilde{\phi}-1}{\Tilde{\phi}}$ for CE-CH pairing. Note that the filling fractions between CE-CE and CE-CH only coincide for $\Tilde{\phi}=2$, so a direct comparison probing particle-holy symmetry can only be made in that case. But the results are still valid for other values. In Fig~\ref{fig:other_filling_fractions}, we show the CE-CE and CE-CH results for $\Tilde{\phi}\in\{4,6\}$. Notably, for both $\Tilde{\phi}=4$ and $\Tilde{\phi}=6$, the favoured channel for the CE-CH theory is still clearly $l=0$, as can be seen in Figs~\ref{fig:cech_4},\ref{fig:cech_6}. For CE-CE pairing however, at $\Tilde{\phi}=4$ $l=1$ is still favoured, although the $l=0$ state (now of CE-CE, not CE-CH) is now significantly closer in strength, as shown in Fig~\ref{fig:cece_4}. At $\Tilde{\phi}=6$ however, the favoured channel for CE-CE looks likely to be $l=0$ and not $l=1$, Fig~\ref{fig:cece_6}. A similar result was seen at low frequencies by \cite{Fu} within the RPA, but they were uncertain as to whether $l=1$ might still be favoured, as it was the more strongly coupled channel at higher frequency. Our MRPA results show that the $l=1$ state might not be favoured significantly at higher frequencies, suggesting that $l=0$ is indeed the most stable pairing for the $\nu=\frac{1}{6}+\frac{1}{6}$ CE-CE theory. It is worth noting that for interlayer paired states the quantized Hall drag resistivity is determined entirely by the pairing channel $l$.  For CE-CH $l=0$ pairing or CE-CE $l=1$ pairing,  one obtains $h/e^2$ quantized Hall drag resistance.   However,  for CE-CE pairing, the case of $l=0$ corresponds to zero Hall drag at zero temperature\cite{SenthilMarstonFisher,ReadGreen,KimNayak}.

\section{Comparison of $l=0$ CE-CH and $l=1$ CE-CE and a possible adjustment to the MRPA}
\label{sec:CECHresults_analytic}
In this section, we first look at the high-frequency limit in both the RPA and MRPA. We find that at finite $m_b$ and large $\omega_m$, there is always a $\log q_c$-divergent, $\mathcal{O}(\omega_m^0)$ difference between the CE-CE and CE-CH Eliashberg anomolous coupling constants. But we find that in the $m_b=0$ MRPA, the difference is non-divergent and only $\mathcal{O}(\omega_m^{-4})$ at large $\omega_m$. We also derive an explicit formula for the difference in anomolous coupling strength between $l=0$ CE-CH and $l=1$ CE-CE, which demonstrates why the MRPA is expected to perform better in the low-frequency limit too. This formula also allows us to propose a possible adjustment to the MRPA, which preserves the correct long-wavelength physics and sum rule behaviour of the MRPA, but guarantees symmetry between the two couplings.
\subsection{A useful formula} \label{sub:diff_formula}
In this section, we take Eq.~\ref{eq:eliashberg_phi} with the MRPA propagator, and compare CE-CH $l=0$ and CE-CE $l=1$. Working through the algebra, we find that
\begin{equation} \label{eq:eliashberg_difference}
    \begin{split}
            \Delta\lambda_{\phi}(\omega_m)&=\lambda_{\phi}^{(l=1),\text{CE-CE}}(\omega_m)-\lambda_{\phi}^{(l=0),\text{CE-CH}}(\omega_m)=\\
            =\frac{m^*}{\pi^2k_F}\mathlarger{\int}_0^{2k_F}\text{d}q& \frac{\sqrt{1-\left(\frac{q}{2k_F}\right)^2}V_{12}(q)\left[1/\Pi_{11}(q,i\omega_m)-f_1\right]^2\times\left[g(q,i\omega_m)\right]^2}{\left[\left(1/\Pi_{00}(q,i\omega_m)+f_1\frac{\omega_m^2}{q^2}-V_{11}(q)\right)\left(1/\Pi_{11}(q,i\omega_m)-f_1\right)-\frac{(4\pi)^2}{q^2}\right]^2-\left[V_{12}(q)\left(1/\Pi_{11}(q,i\omega_m)-f_1\right)\right]^2},\\
            &g(q,i\omega_m)=\left(\frac{1}{\Pi_{00}(q,i\omega_m)}+f_1\frac{\omega_m^2}{q^2}+\frac{2\pi}{m^*}\right)
    \end{split},
\end{equation}
with $f_1=\frac{m^*-m_b}{n_e}$. We choose units with $k_F=1$ and $e=1$, which implies $n_e=\frac{1}{4\pi}$.

 We note that if the denominator of the integrand Eq.~\ref{eq:eliashberg_difference} is positive, the integrand is strictly non-negative, implying $\Delta\lambda_\phi(\omega_m)\geq0$, or that CE-CH $l=0$ is always at least as strong as CE-CE $l=1$. While we do not generally prove the denominator to be positive, this can be clearly seen in the limit of large $d$: there, the second term rapidly decreases at $q\gtrsim1/d$ due to the factor of $V_{12}(q)$. At small $q$, the first term scales as $q^{-4}$, while the second term scales only as $q^{-2}$, meaning the first term is always larger at small $q$: so for large enough $d$, we can confidently claim the denominator to be always positive and that $\Delta\lambda_\phi(\omega_m)\geq0$. In practice, numerically, we have not been able to find any set of values of the parameters where the denominator of Eq.~\ref{eq:eliashberg_difference} is not positive.

\subsection{High-frequency limit}
In Eliashberg theory, $q\leq2k_F$ always, so for $\omega_m\gg\epsilon_F$, we have $q/k_F\ll\omega_m/\epsilon_F$, and some of the results of section \ref{sec:CECHdivergences} may be used, namely the forms of Eqs.~\ref{eq:SmallQPI00},~\ref{eq:SmalQPI11}  giving $(1/\Pi_{11} - f_1) \approx 4 \pi m_b k_F^2$ and $(1/\Pi_{00} + f_1 \omega_m^2/q^2) \approx -4 \pi m_b k_F^2 \omega_m^2/q^2$. Using Eq.~\ref{eq:eliashberg_difference}, we get at $\omega_m\gg \omega_c=1/m_b$,
\begin{equation}
    \Delta\lambda_{\phi}(\omega_m)=\lambda_{\phi}^{(l=1),\text{CE-CE}}(\omega_m)-\lambda_{\phi}^{(l=0),\text{CE-CH}}(\omega_m)=\frac{1}{2\pi^2 k_F \epsilon_F}\int_{q_c}^{2k_F}\text{d}q V_{12}(q)\sqrt{1-\left(\frac{q}{2k_F}\right)^2}.
\end{equation}
While $m_b$ does not appear explicitly here, we note that this result applies only for finite $m_b$ as it requires $\omega_m \gg 1/m_b$. This is divergent, with $\Delta\lambda_{\phi}(\omega_m)\propto\log q_c$ at small $q_c$. So there exists an $\omega_m$-independent, $\log q_c$ divergent difference between CE-CE $l=1$ and CE-CH $l=0$ when $\omega_m\gg \omega_c=1/m_b$. This result is very similar to what we found in section \ref{sec:CECHdivergences}, and using Eq.~\ref{eq:low_q_lambda_phi} we can see that the most divergent term in the coupling is opposite between the two layers, meaning that for small $q_c$ and large $\omega_m$, $\lambda_{\phi}^{\text{CE-CE}, (l=1)}(\omega_m)>0$ and $\lambda_{\phi}^{\text{CE-CH}, (l=0)}(\omega_m)<0$. This is seen for high $\omega_m$ in Fig.~\ref{fig:slager_cutoff_dependance}.

Perhaps the more interesting case is MRPA when $m_b=0$, where all the divergences go away. In order to get a nonzero result, we expand the polarization functions to second lowest order in $\frac{q^2}{\omega_m^2}$, yielding $1/\Pi_{00}\approx-\frac{2\pi}{\epsilon_F}\frac{\omega_m^2}{q^2}\left(1+3\epsilon_F^2\frac{q^2}{\omega_m^2}\right)$ and $1/\Pi_{11}\approx\frac{2\pi}{\epsilon_F}\left(1-\epsilon_F^2\frac{q^2}{\omega_m^2}\right)$. Inserting this into Eq.~\ref{eq:eliashberg_difference} and keeping terms down to $\left(\frac{\epsilon_F}{\omega_m}\right)^4$, we get
\begin{equation} \label{eq:MRPA_CE-CH_diff_high_w_zero_mb}
    \Delta\lambda_{\phi}(\omega_m)\approx \frac{1}{\omega_m^4}\frac{\epsilon_F^3}{32k_F^7}\int_0^{2k_F}\text{d}q V_{12}(q) q^6 \sqrt{1-\left(\frac{q}{2k_F}\right)^2}.
\end{equation}
This is no longer divergent, and scales as $\omega_m^{-4}$, as opposed to the  RPA and finite $m_b$ cases being divergent and scaling as $\omega_m^0$. This shows that in the high-frequency limit, the MRPA $\Delta\lambda_\phi(\omega_m)$ is much smaller than the RPA counterpart --- the MRPA result is much closer to being symmetric between CE-CE and CE-CH pairing than the RPA result is.

\subsection{Low-frequency limit} \label{sub:low_omega_diff}
To find why MRPA would have a lower $\Delta\lambda_\phi(\omega_m)$ at low frequencies, we look at the structure of Eq.~\ref{eq:eliashberg_difference}: the integrand may be written as $(\text{prefactor})\times \left[g(q,i\omega_m)\right]^2$. 
 As mentioned above, we observe that the prefactor is strictly positive for a large range of $q$, $d$ and $\omega_m$.
We thus integrate a positive function, multiplied by $\left[g(q,i\omega_m)\right]^2$, which is itself positive. Thus reducing $\left[g(q,i\omega_m)\right]^2$ reduces the difference between CE-CE and CE-CH pairing strengths.

Since $V_{12}$ becomes very small for $q\gtrsim 1/d$, in the large-$d$ case at least, most of what contributes to $\Delta\lambda_\phi$ is $\left[g(q,i\omega_m)\right]^2$ at small $q$. We now show that at small $q$ and low frequency, the MRPA has a lower $\left[g(q,i\omega_m)\right]^2$. Expanding $g(q,i\omega_m)$ in powers of $\omega_m$ at small $q$, we get
\begin{align}
    \frac{1}{\Pi_{00}(q,i\omega_m)}&\approx-\frac{2\pi}{m^*}-\frac{4\pi m^*}{k_F^2}\frac{\omega_m^2}{q^2}+\mathcal{O}(\omega_m^4),\\
    g(q,i\omega_m)&\approx \left(f_1-\frac{4\pi m^*}{k_F^2}\right)\frac{\omega_m^2}{q^2}+\mathcal{O}(\omega_m^4)= -\frac{4\pi m_b}{k_F^2}\frac{\omega_m^2}{q^2}+\mathcal{O}(\omega_m^4).
\end{align}
Importantly, in RPA ($m_b=m^*$) or in finite $m_b$ MRPA, the leading term is $\mathcal{O}(\omega_m^2)$, while for the $m_b=0$ MRPA the leading term is $\mathcal{O}(\omega_m^4)$. The MRPA cancels the lowest-order term in the Taylor expansion of $g$, reducing it's value at small $\omega_m$ and $q$, which we expect, as discussed, to reduce $\Delta\lambda_{\phi}(\omega_m)$ at least for low $\omega_m$ and large $d$. 

In the last two sections we saw that the MRPA is a step closer to achieving particle-hole symmetry in both low and high frequency limits. As shown in section \ref{sec:CECEresults_numeric}, numeric integration shows it is closer at all $\omega_m$. But there remains a finite difference between $l=0$ CE-CH and $l=1$ CE-CE. In the following, we discuss how the MRPA can be adjusted to achieve perfect degeneracy between the two pairing channels.

\subsection{A possible adjustment to the MRPA}

\label{sub:adjustment}

A hint on how to achieve full symmetry between CE-CE and CE-CH pairing can be seen in Eq.~\ref{eq:eliashberg_difference}: if we can force $g(q,i\omega_m)=0$ for all $q, \omega_m$, the pairing would be exactly symmetric. This would amount to replacing the $\mathcal{F}_1$ of Eq.~\ref{eq:MRPA_F1_def} with an Adjusted MRPA  (AMRPA) version
\begin{equation} \label{eq:AMRPA_F1_def}
    \mathcal{F}^\text{AMRPA}_{1,\mu\nu}(q,i\omega_m)=\left(1-\frac{m_b}{m^*}\right)\begin{pmatrix}
        -\frac{2\pi}{m^*}-\frac{1}{\Pi_{00}(q,i\omega_m)} & 0 \\
        0 & -4\pi k_F^2 m^*  + b(q,i \omega_m)
    \end{pmatrix}_{\mu\nu},
\end{equation}
which is then to be used in Eq.~\ref{eq:K_MRPA_def}. If $m_b=0$, this enforces $g(q,i\omega_m)=0$.
Here the function $b(q,i \omega_m)$ is an arbitrary function, subject to the constraints of causality (no poles in the upper half plane of $b(q,\omega))$), as well as $b(q,\omega) \sim q^n$ for small $q$ with $n \geq 2$ and $b(q,\omega)$ should not grow as fast as $\omega$ for large $\omega$.   Note that for $b=0$ the lower right element of the ${\cal F}_1$ matrix matches that of the MRPA.   The MRPA $\mathcal{F}_1$ can be thought of as the AMRPA $\mathcal{F}_1$, expanded to the lowest order in $q$. As such, the two share the same long-wavelength physics, and the AMRPA will satisfy Kohn's theorem.  It is also easy to show that the AMRPA satisfies the f-sum rule.  

While the AMRPA is designed to give an exact agreement between the coupling strengths of $l=0$ CE-CH and $l=1$ CE-CE, it generalises nicely to a general $l$: within the AMRPA with $m_b=0$, numeric integration shows that in general $l$-wave CE-CH is exactly of the same strength as $(l+1)$-wave CE-CE, at all frequencies. Since CE-CH is symmetric between $l$ and $-l$, this implies that couplings come in degenerate quadruples: $l$ and $(-l)$-wave CE-CH, and $(l+1)$ and $(-l+1)$-wave CE-CE. We note that the AMRPA is a correction within \textit{one} layer which brings the different \textit{bilayer} pairing channels to degeneracy.

\section{The Effect of Imbalance}
\label{sec:imbalance}

R\"uegg, Chaudhary, and Slager\cite{Slager} have also claimed that CE-CE pairing and CE-CH pairing should be inequivalent because they appear to differ when the layers are imbalanced with the total density remaining $\nu_T=1$.  They argue that for CE-CE pairing the two CE Fermi surfaces would become mismatched in size when electrons are transferred from one layer to another, whereas for  CE-CH pairing the two Fermi  surfaces remain the same size.  We disagree with this reasoning: as  mentioned by Wagner et al.\cite{Wagner} (Supplementary material), both cases are equally suited to pairing in the imbalanced case.   For completeness we reiterate this argument here.

We keep the total filling $\nu_T = 1$ fixed, but move some density from one layer to the other so that we have $\nu = \nu_1 +\nu_2$ with $\nu_1 \neq \nu_2$.   For simplicity, let us consider the case where $\nu_1 = p/(2p+1)$, potentially with $p$ large.  This single layer can be described as $p$ filled CE Landau levels.  The opposite layer has filling $\nu_2=(p+1)/(2p+1)$ but this can also be viewed as a filling of $1-\nu_2 = p/(2p+1)$ of holes, and can be described as $p$ filled CH Landau levels.  To build a CE-CH pairing we simply pair the states of the $n^{th}$ CE Landau level of the first layer with those of the $n^{th}$ CH Landau level of the second layer\cite{Wagner}.  However, the second layer can also be thought of as $p+1$ filled Landau levels of 
CEs in a negative magnetic field.   While the description of the $p$ filled Landau levels of CHs is different from that of $p+1$ filled Landau levels of CEs, these two descriptions make almost identical predictions\cite{CooperHalperin,MollerSimonNegativeField,Son}. In particular, the number of states per Landau level is the same and the low energy excitations are in one-to-one correspondence. So while the Fermi seas are different in size, the Fermi surfaces have the same shape, allowing for superconductive pairing.     To build a CE-CE pairing in this picture, one pairs the states of the $(p+1)$th Landau level of CEs in the second layer with the states of the $p$th Landau level of CEs in the first layer. As we show below, this is in fact necessary in order to obtain an $l=1$ BCS state. This allows a perfectly good BCS paired state, except that the lowest Landau level of CEs in the second layer remains unpaired.   However, since this Landau level is buried far below the Fermi surface, it can be considered to be completely filled without any energetic penalty.

We note that this argument does not rely on an integer number of CF LL's being filled: for a general filling fraction of CE's $\nu_\uparrow>\frac{1}{2}$ in the upper layer, $p_\uparrow=\frac{\nu_\uparrow}{2\nu_\uparrow-1}$ CF LL's are filled. If $p_\uparrow$ is non-integer, this indicates partial filling of the top LL. The CE filling fraction in the lower layer is $\nu_\downarrow=1-\nu_\uparrow<\frac{1}{2}$, and $p_\downarrow=\frac{\nu_\downarrow}{1-2\nu\downarrow}=\frac{1-\nu_\uparrow}{2\nu_\uparrow-1}=p_\uparrow-1$. So in the general case, there is an extra Landau level in the upper layer: if the filling fraction is non-integer, both the top LL's have the same fractional filling, meaning they may still be BCS-coupled to form an $l=1$ state.
\subsection{Pairing symmetry in the imbalanced case} \label{sec:wavefunction_symmetry_calc}
In this section, we prove generally that when Landau levels $n$ and $n+l$ are BCS-paired in the imbalanced $\nu_T=1$ case, the resulting pair wave function always has $l$-wave symmetry, both for CE-CE and CE-CH pairing. We take $\nu_\uparrow=\frac{p+1}{2p+1}$, $\nu_\downarrow=\frac{p}{2p+1}$. If the physical magnetic field is $B_0$, the effective fields seen are then exactly opposite between the two layers, defining $B=\frac{1}{2p+1}B_0$, $B_\uparrow=-\frac{1}{2p+1}B_0=-B$ and $B_\downarrow=\frac{1}{2p+1}B_0=B$. From now on we work in units where the effective CF magnetic length is $l_B=\sqrt{\frac{\hbar}{eB}}=1$. Then, in a planar geometry, the single-particle wavefunctions are in the Landau gauge,
\begin{align}
    \Psi_{n,k}^\uparrow(\bold{r})&=e^{iky}\phi_n(x+k) \label{eq:up_WF}\\
    \Psi_{n,k}^\downarrow(\bold{r})&=e^{iky}\phi_n(x-k) \label{eq:down_WF}\\
    \phi_n(x)&=\frac{1}{\pi^{1/4}\sqrt{2^n n!}}e^{-x^2/2}H_n(x), \label{eq:Landau_WF}
\end{align}
with $H_n(x)$ the $n$-th order physicist's Hermite polynomial.

\vspace*{10pt}

\noindent {\bf Case 1: CE-CE pairing}\\
BCS pairing occurs between an electron of momentum $k$ in LL $n+l$ of one layer and an electron of momentum $-k$ in LL $n$ in the other, for all $n$. So the BCS pair function may be constructed as 
\begin{align} \label{eq:BCS_pairfunct_CECE}
    g_l(\bold{r}^\uparrow,\bold{r}^\downarrow)&=\sum_{n=0}^{p-1} g_n f_l^n(\bold{r}^\uparrow,\bold{r}^\downarrow),\\
    f_l^n(\bold{r}^\uparrow,\bold{r}^\downarrow)&=\int\text{d}k\Psi_{n+l,k}^\uparrow(\bold{r}^\uparrow)\Psi_{n,-k}^\downarrow(\bold{r}^\downarrow). \label{eq:BCS_pair_f}
\end{align}
Here, $g_n$ are constants to be determined from energetic considerations. The function $f_l^n$ is the main object of interest and using Eqs.~\ref{eq:up_WF},~\ref{eq:down_WF} we may write it as 
\begin{equation} \label{eq:BCS_f_funct}
    f_l^n(\bold{r}^\uparrow,\bold{r}^\downarrow)=\int\text{d}k e^{ik\left(y^\uparrow-y^\downarrow\right)}\phi_{n+l}(x^\uparrow+k)\phi_{n}(x^\downarrow+k)
\end{equation}

\vspace*{10pt}

\noindent {\bf Case 2: CE-CH pairing}\\
Here, both layers are less than half-full of the carriers, so the wavefunction Eq~\ref{eq:down_WF} is to be used for both. Now we pair a hole of momentum $k$ in one layer to an electron of momentum $k$ in the other. Taking the hole to be in the up layer, making the replacement $k\to -k$ in Eq.~\ref{eq:down_WF} and remembering that the hole wavefunction must be complex conjugated we get, since $\phi_n$ is real,
\begin{align}
    g_l(\bold{r}^\uparrow,\bold{r}^\downarrow)&=\sum_{n=0}^p g_n f_l^n(\bold{r}^\uparrow,\bold{r}^\downarrow),\\
    f_l^n(\bold{r}^\uparrow,\bold{r}^\downarrow)&=\int\text{d}k e^{ik\left(y^\uparrow-y^\downarrow\right)}\phi_{n+l}(x^\uparrow+k)\phi_{n}(x^\downarrow+k).
\end{align}
This exactly matches the expression Eqs.~\ref{eq:BCS_pairfunct_CECE},~\ref{eq:BCS_f_funct}, for CE-CE, so both may be treated simultaneously from now on.

\vspace*{10pt}

To evaluate $f_l^n$ explicitly, we define $\bold{R}=\frac{1}{2}\left(\bold{r}^\uparrow+\bold{r}^\downarrow\right)$ and $\bold{r}=\bold{r}^\uparrow-\bold{r}^\downarrow$. Using these in Eq.~\ref{eq:BCS_f_funct}, and putting $q=k-X+x/2$, we get 
\begin{equation}
    f_l^n(\bold{R},\bold{r})=e^{i\left(\frac{x}{2}-X\right)y}\int\text{d}q e^{iqy}\phi_{n+l}(x+q)\phi_n(q)
\end{equation}
Using Eq.~\ref{eq:Landau_WF}, we can get 
\begin{align}
    f_l^n(\bold{R},\bold{r})&=C_l^ne^{i\left(\frac{x}{2}-X\right)y}\int\text{d}q e^{iqy}e^{-\frac{(x+q)^2}{2}-\frac{q^2}{2}}H_{n+l}(x+q)H_n(q)   \label{eq:f_no_more_fourier0}\\
    f_l^n(\bold{R},\bold{r})&=C_l^ne^{-iXy}e^{-\frac{1}{4}\left(x^2+y^2\right)}\int\text{d}k e^{-k^2}H_{n+l}(k+\frac{x+iy}{2})H_n(k-\frac{x-iy}{2}) \label{eq:f_no_more_fourier}
\end{align}
Here, $C_l^n=\frac{1}{\sqrt{\pi 2^{2n+l} n! (n+l)!}}$ is a constant. Going between Eq.~\ref{eq:f_no_more_fourier0} and Eq.~\ref{eq:f_no_more_fourier}, we re-wrote the exponential as $iqy-\frac{(x+q)^2}{2}-\frac{q^2}{2}=-(q+\frac{x-iy}{2})^2-\frac{1}{4}\left({x^2+y^2+2ixy}\right)$. We then defined $k=q+\frac{x-iy}{2}$. 
Note the identity
\begin{equation}
    \int_{-\infty}^{\infty} \text{d}x~e^{-x^2} H_a(x+y)H_b(x+z)=\sqrt{\pi}2^ab!~y^{a-b}L_b^{(a-b)}(-2yz)
\end{equation}
with $L_n^{(\alpha)}$ associated Laguerre polynomials and $a\geq b$. Using this on Eq~\ref{eq:f_no_more_fourier} yields
\begin{align}
    l\geq0:\hspace{10px }f_l^n(\bold{R},\bold{r})&=e^{-iXy}(x+iy)^l\sqrt{\frac{n!}{2^l(n+l)!}}e^{-\frac{1}{4}|\bold{r}|^2}L_n^{(l)}\left(\frac{1}{2}|\bold{r}|^2\right) \label{eq:f_final_form}\\
    l<0:\hspace{10px }f_l^n(\bold{R},\bold{r})&=e^{-iXy}(x-iy)^{|l|}\sqrt{\frac{n!}{2^{|l|}(n+|l|)!}}e^{-\frac{1}{4}|\bold{r}|^2}L_n^{(|l|)}\left(\frac{1}{2}|\bold{r}|^2\right) \nonumber
\end{align}
The prefactor $e^{-i X y}$ in Eq.~\ref{eq:f_final_form} is a center of mass momentum in the $X$ direction depending on the separation in the $y$ direction.  This stems from the non-commutativity of $x$ and $y$ coordinates in a magnetic field, and is reminiscent of the physics of excitons in a magnetic field  (see for example Ref.~\cite{Kallin}).

Focusing on $l\geq0$ for now, the remainder of  Eq.~\ref{eq:f_final_form}  has the form $(x+iy)^l$, multiplied by a function of only $|\bold{r}|$, characteristic of an $l$-wave state. Of course the final pair wavefunction is $g_l(\bold{r})$. All the angular dependencies factor out, and we get from Eqs.~\ref{eq:BCS_pairfunct_CECE},\ref{eq:f_final_form}
\begin{align} \label{eq:g_final_form}
    g_l(\bold{R},\bold{r})&=(x+iy)^le^{-iXy}G_l(|\bold{r}|^2) \\
    G_l(|\bold{r}|^2)&=e^{-\frac{1}{4}|\bold{r}|^2}\sum_{n=0}^{p-1}g_n\sqrt{\frac{n!}{2^l(n+l)!}} L_n^{(l)}\left(\frac{1}{2}|\bold{r}|^2\right),\nonumber
\end{align}
clearly showing an $l$-wave pair wavefunction for $l\geq0$. We note that $G_l(|\bold{r}|^2)$ is just a polynomial of order $p-1$ in $|\bold{r}|^2$, multiplied by the Gaussian factor $e^{-\frac{1}{4}|\bold{r}|}$. Trivially, for $l<0$, we get $g_l(\bold{R},\bold{r})=e^{-iXy}(x-iy)^{|l|}G_{|l|}(|\bold{r}|^2)$, which gives a $-|l|=l$-wave state, as expected.

To re-iterate our argument, in the case of CE-CH pairing, there are $p$ filled Landau levels in each layer: we may pair levels $n$ and $n$ for $0\leq n \leq p-1$, giving us an $l=0$ BCS state. In the case of CE-CE pairing, there are $p+1$ LLs in the upper layer and $p$ in the lower. We pair level $n+1$ in the upper layer to level $n$ in the lower layer for $0\leq n \leq p-1$. This results in an $l=1$ BCS state, and only the LLL of the upper layer remains unpaired. This level is far below the Fermi surface, and as such the fact that it remains unpaired should not significantly impact the pairing strength.

\subsection{Weak field limit}

Consider the limit of our wavefunction as $\nu\to\frac{1}{2}$. Re-introduce factors of $l_B$ into Eq.~\ref{eq:f_final_form}, and take $l_B\to\infty$ (this is the CF $l_B$). Measuring distance in units of the original magnetic length, $l_{B_0}$, we get, with $s=\text{sign}~l$
\begin{equation} \label{eq:f_weak_field_limit}
    f_l^\eta(\bold{r})=\left(\frac{x+isy}{|\bold{r}|}\right)^{|l|} J_{|l|}\left(\sqrt{\eta}|\bold{r}|\right),
\end{equation}
where $\eta$ replaces $n$, being defined as $\eta=n\frac{\omega_{c,\text{CF}}}{\epsilon_F}$, in other words $\eta=\frac{\epsilon}{\epsilon_F}$ is the ratio of the energy of the electron under consideration to the Fermi energy of the $\nu=\frac{1}{2}$ system. As usual, $J_{|l|}$ is the Bessel function of the first kind and is of order $|l|$.

Importantly, as discussed above, if $\nu_\text{tot}=1$, we will \textit{always} have $l=1$ for CE-CE and $l=0$ for CE-CH, if we want to pair the states closest to the Fermi surface, at least at finite imbalance. We expect this to continue to hold into the limit of infinitesimal imbalance.
The full BCS pair wavefunction is then of the Hankel form:
\begin{equation} \label{eq:g_weak_field_limit}
    g_l(\bold{r})=\left(\frac{x+isy}{|\bold{r}|}\right)^{|l|} \int\text{d}\omega g(\omega) J_{|l|}\left(\sqrt{\frac{\omega}{\epsilon_F}}|\bold{r}|\right) =  e^{i \theta l} \,\,  \tilde g(|{\bf r}|),
\end{equation}
where again $g(\omega)$ is to be determined by energetic considerations, which can then generate any function $\tilde g$ whose Taylor expansion has the leading power $|{\bf r}|^{|l|}$.  If we started by considering plane waves and insisted the pairing to be $l$-wave, the resulting pair wavefunction in two dimensions is exactly Eq.~\ref{eq:g_weak_field_limit}. This is another way to understand why $l=0$ CE-CH and $l=1$ CE-CE would be preferred: they are the natural continuation of the finite magnetic field pair function into the zero-field limit.

\subsection{Anisotropic case} \label{sec:imbalance_anisotropic}
One might ask about the stability of the above wavefunction approach against slight anisotropic deformations. Consider an anisotropic effective band Hamiltonian,
\begin{equation} \label{eq:anisotropic_imbalance}
    H = \frac{1}{2m}\left[a\Pi_x^2+\Pi_y^2/a\right]=\frac{1}{2m}\left[-a\frac{\partial^2}{\partial x^2}+\frac{1}{a}\left(-i\frac{\partial}{\partial y} - Bx \right)^2\right]
\end{equation}
where $\bold{\Pi}=\bold{p}-\bold{A}$. We work in the Landau gauge, $\bold{A}=(0,Bx)$. Define $\Tilde{x}=x/\sqrt{a}$ and $\Tilde{y}=\sqrt{a}y$, along with $\Tilde{\bold{p}}=(-i\partial_{\Tilde{x}},-i\partial_{\Tilde{y}})$. In these transformed coordinates, the Hamiltonian reads\cite{KunYangAnisotropy}
\begin{equation}
    H=\frac{1}{2m}\left[-\frac{\partial^2}{\partial \Tilde{x}^2}+\left(-i\frac{\partial}{\partial \Tilde{y}} - B\Tilde{x} \right)^2\right] =\frac{1}{2m}\left[\Tilde{\Pi}_{\Tilde{x}}^2+ \Tilde{\Pi}_{\Tilde{y}}^2\right],
\end{equation}
of course $\bold{\Tilde{\Pi}}=\bold{\Tilde{p}}-\bold{\Tilde{A}}$ and $\bold{\Tilde{A}}=(0,B\Tilde{x})$. This is just the isotropic Hamiltonian we have already analysed in Sect.\ref{sec:wavefunction_symmetry_calc}. Note that since the product of $x$ and $y$ is preserved, so are areas and thus filling fractions do not change (crucially, the magnetic field $B$ is the same in both equations). We thus expect all our analysis to carry over to this case. There is a difference however, namely that if the potential $V(\bold{r})$ was isotropic in the original $(x,y)$ coordinates, it will no longer be isotropic in our rescaled $(\Tilde{x},\Tilde{y})$ coordinates. But note that we never appealed to the pairing interaction being isotropic -- we merely relied on the assumption that the most strongly coupled states are the ones where particles closest to the Fermi surface are all paired up. If this was true in the isotropic case, there is no obvious reason for why it would no longer hold in the anisotropic case (as all that really matters is that the effective interaction is strongest at low frequency, which must be true at least close to the critical temperature), and we conclude that our analysis of the isotropic imbalanced state carries over completely to the anisotropic case, still favouring $s$-wave pairing for CE-CH and $p$-wave ($l=1$) for CE-CE in the imbalanced case   and as discussed, also in the limit $\nu\to \frac{1}{2}$.  Note that the angle-dependent part of the wavefunction is now $\sim(\Tilde{x}+i\Tilde{y})^l$. In the original coordinates the angle dependent part of the pairing translates to $\sim(x/\sqrt{a}+i\sqrt{a}y)^l$, so it is $l$ wave pairing only in the rescaled coordinates (and similarly we should use $|\bold{r}|=\sqrt{x^2/a + ay^2 }$ in the wavefunction formulae).    Indeed, once we have broken rotational symmetry the pairing symmetry need not be an eigenstate of rotation (i.e., have a fixed $l$) in the unrescaled coordinate system.

If we instead made the potential anisotropic and kept the band structure isotropic (or if we made both anisotropic), we would expect a similar result to continue to hold --- again the symmetry of the paired wavefunction does not primarily come from properties of the interaction between particles, but rather from the Landau Level structure of the problem at finite imbalance. It can be shown that the composite fermion Landau level structure of incompressible FQHE states is stable against slight anisotropy in the form of a tilted magnetic field \cite{Yang_tilded_field_charged,Yang_tilded_field_neutral}. The effects of a tilted field on those states are similar to the effects on the Laughlin wavefunction \cite{Papic_Tilted_Field}. 

In addition, as mentioned in the main text, the arguments of Ref.~\cite{KunYangAnisotropy} show that any system with Gaussian interaction is completely stable against anisotropy.   In that work the focus was on the $\nu=\frac{1}{2}$ state where anisotropy was explored in experiment\cite{PhysRevLett.110.206801}.  We have examined the Gaussian interaction in our CSMRPAE calculations and found the same key predictions (CE-CH pairing in $l=0$ almost degenerate with CE-CE pairing in $l=1$), as shown in Sect.~\ref{sub:potentials}.  At least in this case we can then conclude that we are extremely robust against geometric deformation.

While strong anisotropy, especially in higher Landau levels might lead to stripe ordering \cite{Papic_Haldane,PhysRevB.86.035122}, for the lowest Landau level one needs strong anisotropy to favor this. The arguments presented here lead us believe that our results (that $l=1$ is the favoured channel for CE-CE pairing and that $l=0$ is favoured for CE-CH pairing and that these two channels are very close in strength) should be stable against a slight anisotropy in the system.

\section{Robustness of near degeneracy and particle-hole symmetry}

It is natural to ask about the relation of particle-hole symmetry of the LLL and the (near) degeneracy of  p-wave CE-CE pairing and the s-wave CE-CH pairing. 
In the Dirac composite-fermion language, Sodemann et al\cite{Sodemann} write a pairing term $i \Delta \psi \sigma_y \tau_x \psi + h.c.$ where $\psi$ is a 4-spinor representing the Dirac fermion in each layer, $\Delta$ is the pairing amplitude, $\tau_x$ exchanges the two layers and $i \sigma_y$ is the CT operator (particle-hole conjugation).   Thus this term pairs particles in one layer with holes in the opposite layer forming excitons.   Sodemann et al argue that this is equivalent to CE-CE p-wave pairing.    Assuming point pairing (s-wave) in the Dirac representation, and $\tau_x$ as the layer exchange, this form is actually unique.  
I.e., we can't pair particles to particles by dropping $i \sigma_y$, or changing it to $\sigma_x$ or $\sigma_z$ because the term would vanish by fermionic symmetry.     This might suggest that translation to the Dirac picture gives a reason why in the HLR picture, CE-CE p-wave and CE-CH s-wave are identical --- there is only one choice of a pairing channel.     (If we add angular momentum to the pairing term explicitly, we change the shift of the wavefunction and this can then be distiguished as a different state.)

While this type of argument is suggestive, it is not conclusive.  The issue is that one could have chosen a different representation $\tau_y$ of the layer exchange operator which then allows different s-wave pairing channels, and we do not, as of yet, see why these might not also be allowed --- thus giving different possible pairing channels in the Dirac representation.   One cannot then appeal to the uniqueness of the pairing term to argue that both types of paired states must be identical.   (Acknowledgement: much of this above discussion comes from recent correspondence with D. X. Nguyen).  

It is also worth asking whether the near degeneracy that we have found between CE-CE p-wave and CE-CH s-wave pairing really relies on particle-hole symmetry.  The underlying system is only particle-hole symmetric in the limit of $m_b \rightarrow 0$.  Further, even in this limit, the HLR approach\cite{HLR} violates particle-hole symmetry once any mean field approximation is made.   Nonetheless, even without the $m_b \rightarrow 0$ limit,  a surprising emergent particle-hole symmetry can arise near half-filling\cite{CooperHalperin}.  In our Eliashberg calculation, particle-hole symmetry is certainly broken at frequency scales above the cyclotron energy --- which (for small enough $m_b$) should not play a role in the low energy physics.   However, at low energy, the (approximate) symmetry seems robust to many variations in details of the calculation.   Nonetheless, one can certainly break this symmetry explicitly at low energy by including explicitly particle-hole breaking terms such as three-body interactions, or appropriately particle-hole breaking disorder.    In the next section, we show that the CE-CE $l=+1$ pairing and CE-CH $l=0$ pairing remain degenerate even in the presence of certain three-body interactions, showing that the symmetry between the different pairing channels does not arise from the particle-hole symmetry of the underlying model.

\subsection{Three-body interactions}
We turn to the effect of three-body interactions on our results. While such interactions break the particle-hole symmetry of the problem, we surprisingly find that for certain types of three-body terms, the resulting degeneracy between the CE-CE $l=1$ and CE-CH $l=0$ channels is preserved.
We assume that the three-body terms of interest may be written as an interaction between three density fluctuations. Starting out in the CE-CE picture, we may write the additional term as $L_\text{3-body}=\sum_{s,t,u\in\{1,2\}} \int\text{d}\bold{x}\text{d}\bold{y}\text{d}\bold{z}W_{stu}(\bold{x},\bold{y},\bold{z})\delta\rho_s(\bold{x})\delta\rho_t(\bold{y})\delta\rho_u(\bold{z})$, with $s,t,u$ layer indices. We assume translational invariance, so that $W_{stu}$ depends only on the differences of the three coordinates. The main idea is to use the flux attachment condition, Eq.~\ref{eq:gauge_curl_electron_density} in the form $\delta\rho_s(\bold{q})=\frac{1}{4\pi}|\bold{q}|\tilde{a}^{(s)}_1(\bold{q})$, with $\tilde{a}^{(s)}_1(\bold{q})$ the fluctuation of the transverse gauge field about its mean-field value. Then, the three-body term turns into a 3-point vertex for the transverse component of the gauge field. This affects our results by modifying the gauge propagator via the loop diagram shown in Eq.~\ref{eq:3-body-feynman}.

\begin{equation} \label{eq:3-body-feynman}
\begin{tikzpicture}
  \begin{feynman}
    \vertex (a);
    \vertex [right=of a] (b);
    \vertex [right=of b] (c);
    \vertex [right=of c] (d);
   \diagram* {
  (a) -- [photon, momentum=\(q\)] (b)
    -- [photon, half left, looseness = 1.5, momentum=\(p+q\)] (c)
    -- [photon, half left, looseness = 1.5, momentum=\(p\)] (b),
  (c) -- [photon, momentum = \(q\)] (d),
    };
\end{feynman}
\end{tikzpicture}
\end{equation}

\subsubsection{Asymmetric between the two layers}
First, consider adding such a three-body term that after exchanging the layer indices, $L_\text{3-body}$ neither stays invariant, nor goes to $-L_\text{3-body}$. In the $a_\mu^{\pm}$ basis defined in Eq.~\ref{eq:gauge_transformation}, swapping the layers corresponds to flipping the sign of $a^{-}_\mu$, so such a term would have to involve both terms with even and terms with odd powers of $a^{-}_\mu$ after doing the flux attachment. With that, it can be shown that the diagram of Eq.~\ref{eq:3-body-feynman} leads to a nonzero $\left\langle a^+_\mu(q,i\omega_m)a^-_\nu(-q,-i\omega_m)\right\rangle$, mixing the in-phase and out-of-phase components of the gauge fluctuations. This may significantly modify the behaviour of the system and could lead to a breaking of the degeneracy between the CE-CE $l=1$ and CE-CH $l=0$ channels, but a full calculation is beyond the scope of the current work.
\subsubsection{Symmetric between the two layers}
The alternative is that upon swapping the layer indices, $L_\text{3-body}\rightarrow \pm L_\text{3-body}$. This implies that after flux attachment, $L_{\text{3-body}}$ contains only even ($+$) or only odd ($-$) powers of $a^{-}$. In such a scenario, $\left\langle a^+_\mu(q,i\omega_m)a^-_\nu(-q,-i\omega_m)\right\rangle=0$ even after the inclusion of the three-body term, so the fluctuations do not mix. Noting that the interaction is only between the $\mu=1$ components of the gauge field, the diagram in Eq.~\ref{eq:3-body-feynman} to lowest order leads to the modification of the gauge propagator
\begin{align}
    \mathcal{D}^{\text{3-body}}_{\pm,\mu\nu}(q,i\omega_m)&=\mathcal{D}_{\pm,\mu\nu}(q,i\omega_m)+q^2 \mathcal{D}_{\pm,\mu1}(q,i\omega_m)\mathcal{D}_{\pm,1\nu}(q,i\omega_m)F^\pm(q,i\omega_m) \label{eq:gauge_threebody_propagator},\\
    F^\pm(q,i\omega_m) &= \sum_{A,B=\pm} \sum_{\omega_l}\int\text{d}^2\bold{p} \Tilde{W}^\pm_{AB}(\bold{p},\bold{q}) |\bold{p+q}||\bold{q}|\mathcal{D}_{A,11}(\bold{q+p},i\omega_m+i\omega_l)\mathcal{D}_{B,11}(-\bold{p},-i\omega_l) \label{eq:loop_integral_gauge}.
\end{align}
Here, the coefficients $\Tilde{W}^\pm_{AB}$ depend on the details of the three-body interaction and are quadratic in the Fourier transforms of the $W_{stu}$ above (specifically, if $W_{stu}$ is short-ranged, then $\Tilde{W}^\pm_{AB}$ will be momentum-independent at low momenta). Computing the loop integral Eq.~\ref{eq:loop_integral_gauge} using the bare gauge propagator gives zero. While we do not evaluate it using the (A)MRPA, we know that due to the lack of IR divergences in $\mathcal{D}$ for $m_b=0$, $F(q,i\omega_m)$ will be finite in those two approximations, assuming a well-behaved $\Tilde{W}^\pm_{AB}$.

Now consider what changes when we particle-hole conjugate one of the layers, say layer $(1)$. The effect is two-fold: first, for each factor of $\delta\rho_1$ in any term of $L_\text{3-body}$, we pick up a factor of $(-1)$ due to Fermi statistics. But second, from Eq.~\ref{eq:gauge_curl_electron_density}, an additional factor $(-1)$ per $\delta\rho_1$ arises due to the opposite flux attachment sign for CH's. So the three-body term, expressed in terms of the gauge fields, is invariant under particle-hole conjugation. 

One might worry that much like PH-conjugating a two-body term leads to an induced one-body term due to normal ordering (Appendix B of \cite{Simon_PH_induced_term}), PH-conjugating a three-body term would induce additional two-body and one-body terms. While this generally happens, note that our term is written out in terms of density fluctuations $\sim \left(\psi^\dag(\bold{x})\psi(\bold{x})-n_e\right)\left(\psi^\dag(\bold{y})\psi(\bold{y})-n_e\right)\left(\psi^\dag(\bold{z})\psi(\bold{z})-n_e\right)$. When multiplied out, this implicitly includes multiple one-body and two-body terms. It can be shown that at $\nu=\frac{1}{2}$, the difference between these terms and their counterparts in the equivalent CH expression exactly cancels the induced two and one-body terms from PH conjugation, meaning that the PH-conjugated term has the same form as the original, multiplied by a factor of $(-1)$ for each PH-conjugated density, as claimed above.

Using the invariance of the three-body term, and thus also the gauge field propagator under PH conjugation of either layer, it can be shown analytically (in a way similar to Sect.~\ref{sec:CECHresults_analytic}) that within the AMRPA, $\Delta\lambda_{\phi}(\omega_m)=\lambda_{\phi}^{(l=1),\text{CE-CE}}(\omega_m)-\lambda_{\phi}^{(l=0),\text{CE-CH}}(\omega_m)=0$ holds with the three-body term for arbitrary finite $F^\pm$ (note that $F^+$ and $F^-$ are generally not equal, but this is not required for the cancellation). This follows from using the tensor structure of the correction $\sim \mathcal{D}_{\mu1}\mathcal{D}_{1\nu}$ in Eq.~\ref{eq:eliashberg_phi}. This suggests that the MRPA, which is very similar to the AMRPA, will see only a small breaking of the degeneracy. While we do not compute the loop integrals $F^\pm$, approximating them as polynomials in $q$ and scanning through a wide range of the coefficients, we find no cases where $\Delta\lambda_\phi(\omega_m)$ is appreciably larger than what is seen in Sect.~\ref{sec:CECEresults_numeric} without a three-body term. We thus conclude that layer-exchange (anti-)symmetric terms cubic in the gauge fields do not break the degeneracy between the CE-CE $l=1$ and CE-CH $l=0$ channels. While they cannot break this degeneracy, three-body terms can significantly impact the pairing strength of any channel. It remains to be seen whether a strong enough three-body term might lead to a different pairing channel being favoured, or if it can entirely prevent pairing.

We thus find that three-body terms which (1) may be written as a product of three density fluctuations and (2) are either fully symmetric or fully antisymmetric under layer exchange in the CE-CE picture maintain the near (full) degeneracy of the CE-CE $l=1$ and CE-CH $l=0$ channels within the MRPA (AMRPA), despite breaking particle-hole symmetry. While we suspect that terms not obeying these two conditions could lift the degeneracy, a more detailed calculation would be required in that case.

\end{document}